# A Geometric Analysis-Based Safety Assessment Framework for Marine Vehicle Route Decision-Making

Zilong Xu[a], Zihao Wang[a,b,*], He Li[c], Dingli Yu[c], Zaili Yang[c], Jin Wang[c]

[a] *Institute of Artificial Intelligence, Shanghai University, 99 Shangda Road, Shanghai, 200444, Shanghai, China*

[b] *Engineering Research Center of Unmanned Intelligent Marine Equipment, Ministry of Education, 99 Shangda Road, Shanghai, 200444, Shanghai, China*

[c] *School of Engineering, Liverpool John Moores University, Byrom Street, Liverpool, L33AF, Merseyside, United Kingdom*

**Abstract:** This paper develops a Geometric Analysis-based Route Safety Assessment (GARSA) framework to enhance the safety of marine vehicles navigating in restricted waters. Utilizing line and point geometric elements to define waterway boundaries, the framework enables the construction of a dynamic width characterization function to quantify spatial safety within complex restricted navigation spaces. An iterative method is developed to calculate the dynamic width characterization function, enabling an abstracted spatial property representation of waterways. Based on this, a navigational safety index that accounts for the overall route safety as well as the spatial constraints of local segments is created, enabling the selection of the safest waterway to pass through. Case studies in Hamburg Port and hypothetical waterways show that GARSA identifies spatial safety differences among routes. For example, GARSA assigned safety assessment values of 0.6 and 135 to two intra-port transit routes of 7,258 m and 7,845 m, respectively, with the higher value indicating a safer route. Overall, GARSA provides a quantitative basis for safety-oriented route decision-making of marine vehicles in restricted waters.

## 1. Introduction

Autonomous Marine Vehicles enable navigation and operations in complex environments without direct human intervention [1, 2], driven by advancements in automation, robotics, multi-sensor perception, and intelligent decision-making concepts [3, 4]. Among diverse operational environments, restricted waters, such as ports and intricate waterways, pose significant safety challenges due to narrow and irregular geometries [5, 6].

Operating in such environments requires safe autonomous navigations [7, 8]. The autonomous control framework typically operates on two hierarchical layers: (i) a decision-making layer, responsible for identifying feasible routes and selecting the optimal route based on performance evaluations [9, 10]; and (ii) a motion control layer, tasked with executing the planned route to achieve precise vessel maneuvering [11, 12].

In typical navigation scenarios in restricted waters, such as entering and exiting ports or traversing complex restricted passages, route decision-making is to understand and mitigate collision risks. Effective route decision enables vessels to avoid excessively narrow or rugged sections, thereby reducing the likelihood of collisions and minimizing transportation disruptions. This task relies heavily on the understanding of irregular geometries of the navigable environment and the assessment of the safety properties of navigational routes. It addresses two key aspects: (i) route network modeling, which identifies navigational pathways between decision points, and (ii) safety assessment assisted route selection, which evaluates routes based on safety criteria and determines the safest one. Route network modeling refers to constructing an abstract representation of such environments to capture the topological structure. In localized environments, this differs significantly from global route network modeling, which leverages abundant Automatic Identification System (AIS) data [13-15]. Localized route network modeling often lacks sufficient navigational data and requires tailored approaches. Common methods involve selecting key nodes and connecting them if the line between two nearest nodes is within feasible regions. For instance, Jang et al. [16] used a random generation method to obtain nodes in feasible regions. However, this method may generate redundant nodes in complex navigation spaces. Zhou et al. [17] connected boundary elements of polygonal obstacles to construct the network, while Shah et al. [18] simplified networks by removing redundant nodes. Additionally, Schoener et al. [19] utilized Voronoi diagrams to divide the navigable domain into regions, deriving central axes as potential pathways. These methods effectively capture connectivity but generally focus on distance attributes, with limited integration of safety metrics, especially in geometrically complex environments. To enhance safety assessment, integrating safety-related evaluation metrics into the network model is essential.

In restricted waters, assessing navigational safety for ships presents unique challenges. The limited space reduces the margin for maneuvering errors, demanding a more accurate representation of risk factors. Bayesian network methods [20-22] typically represent risk levels in a non-quantitative manner, which limit their ability to accurately capture the diversity of risk factors. In contrast, risk index methods allow for a more precise quantification of risk by incorporating parameters such as ship speed [23], ship-to-shore distance [24], distance at the closest point of approach (DCPA) [25], and time to the closest point of approach (TCPA) [26]. However, risk index methods primarily focus on localized risks between vessels and nearby obstacles and provide limited representation of the overall geometric variations of irregular waterways. This limitation makes it difficult to accurately capture the uneven distribution of spatial safety within such environments.

Existing route decision-making methods in restricted waters generally focus on improving navigational efficiency, such as optimizing route length and travel time [28], while navigational safety is often treated as a secondary consideration or being neglected. More critically, existing studies lack a mathematical function that maps the irregular geometry of waterways into a quantitative representation of spatial safety. Therefore, it is essential to incorporate a safety assessment mechanism into the route decision-making process.

To prioritize safety in route decision-making for irregularly shaped waterways, this paper aims to develop a geometric analysis-based route safety assessment (GARSA) framework that can address both route exploration and quantitative safety assessment. The novel contribution concludes:

- Propose a dynamic width characterization method to capture spatial variations in irregular waterways, providing inputs for navigational safety assessment.
- Design an iterative algorithm to construct the dynamic width characterization function, enabling the abstract representation of environmental spatial factors and providing a practical approach to environmental state awareness.
- Develop a safety index to balance global navigational safety and local risk for safe route determination.

The efficiency and feasibility of GARSA are validated by the estimation of Hamburg Port, the results confirmed that the method provides decision support for safe route selection in complex restricted environments. Overall, the proposed safety assessment method supports geometric feature qualification and safety control of automatic devices.

The structure of this paper is as follows. Section 2 introduces the dynamic width characterization function. Section 3 addresses the safety route decision-making. Section 4 presents results validation. Section 5 concludes the paper.

## 2. Dynamic Width Characterization Function

The navigational safety of a regularly shaped waterway can be assessed based on width, which reflects the potential risk of collision with shore. However, irregular boundaries of waterways endure variable width at different locations/points. To assess the navigational safety of an entire irregularly shaped waterway, a dynamic width characterization function is needed to capture the variation in waterway.

### 2.1. Dynamic Width Characteristic

An irregularly shaped waterway can be viewed as consisting of multiple connected sub-waterway units. The boundaries of these units can be further simplified and represented by point and line features $H_i$ $(i=1,2,\dots,n)$ with each waterway composed of three connected sub-waterway units, see Figure 1. The sub-waterway units can be approximated using combinations of 'line-line', 'point-line', or 'point-point' arrangements. On this basis, the safety assessment of the waterway can be conducted within each sub-waterway unit.

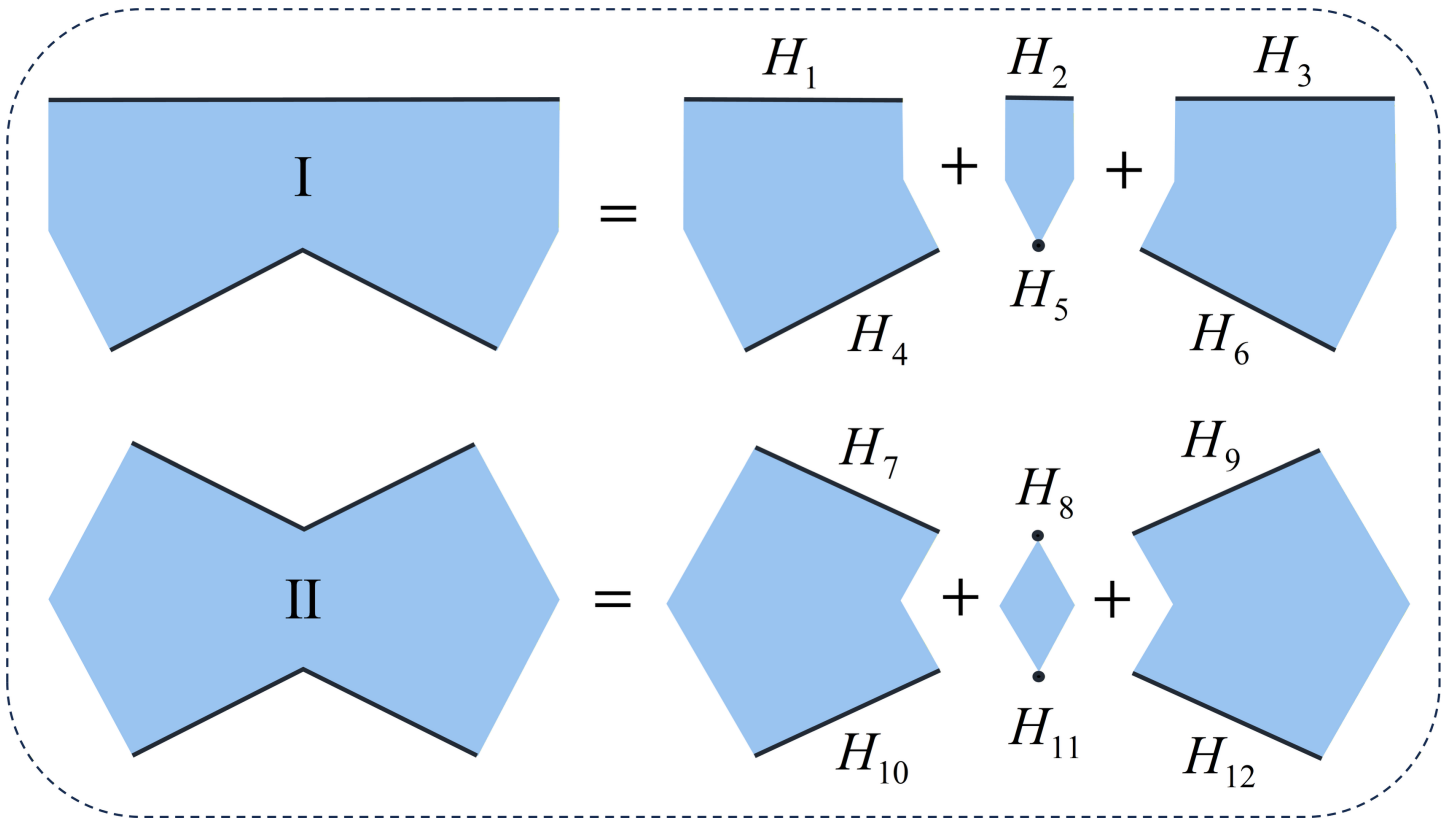

Figure. 1. Schematic of irregular waterways using point and line elements.

**Definition 1** (*Dynamic width characteristic*). Given the point and line features $H_i$ $(i=1,2,\dots,n)$ extracted from the waterway boundary, let $L$ denotes a cross-section of the navigable space at a specific position along the waterway. The cross-section is generated according to the local boundary geometry in the calculation process (see Section 2.2.). For any point $P_j \in L$, the minimum distance from $P_j$ to all surrounding geometric elements $H_i$ is denoted as $w_{P_j}$, say:

$$w_{P_j} = \min_{1\le i\le n} dist(P_j, H_i), \tag{1}$$

where $dist(P_j, H_i)$ represents the distance between $P_j$ and $H_i$. Then, $\exists P_q \in L$ such that:

$$w_{P_q} \ge w_{P_j} \quad \text{s.t.}\, P_j \in L, \tag{2}$$

where $w_{P_q}$ is the maximum value among the local minimum distances $w_{P_j}$ along the cross-section $L$, and $\omega_{dwc} := w_{P_q}$ is defined as the dynamic width characteristic at this location.

The dynamic width characteristic is illustrated with an irregularly shaped waterway as shown in Figure 2. The waterway boundary can be roughly decomposed into three-line features, denoted as $H_1$, $H_2$, and $H_3$, and cross-sections $L_a$, $L_b$, and $L_c$. Hence, there exists a point $\mathrm{P}_d$ within $L_a$, where the minimum distance to the surrounding geometric elements, denoted as $w_{P_d}$, is greater than that at any other point within $L_a$. Therefore, the dynamic width characteristics of $L_a$ can be determined by $\mathrm{P}_d$ as $w_{P_d}$. Similarly, the dynamic width characteristic of cross-sections $L_b$ and $L_c$ can be determined.

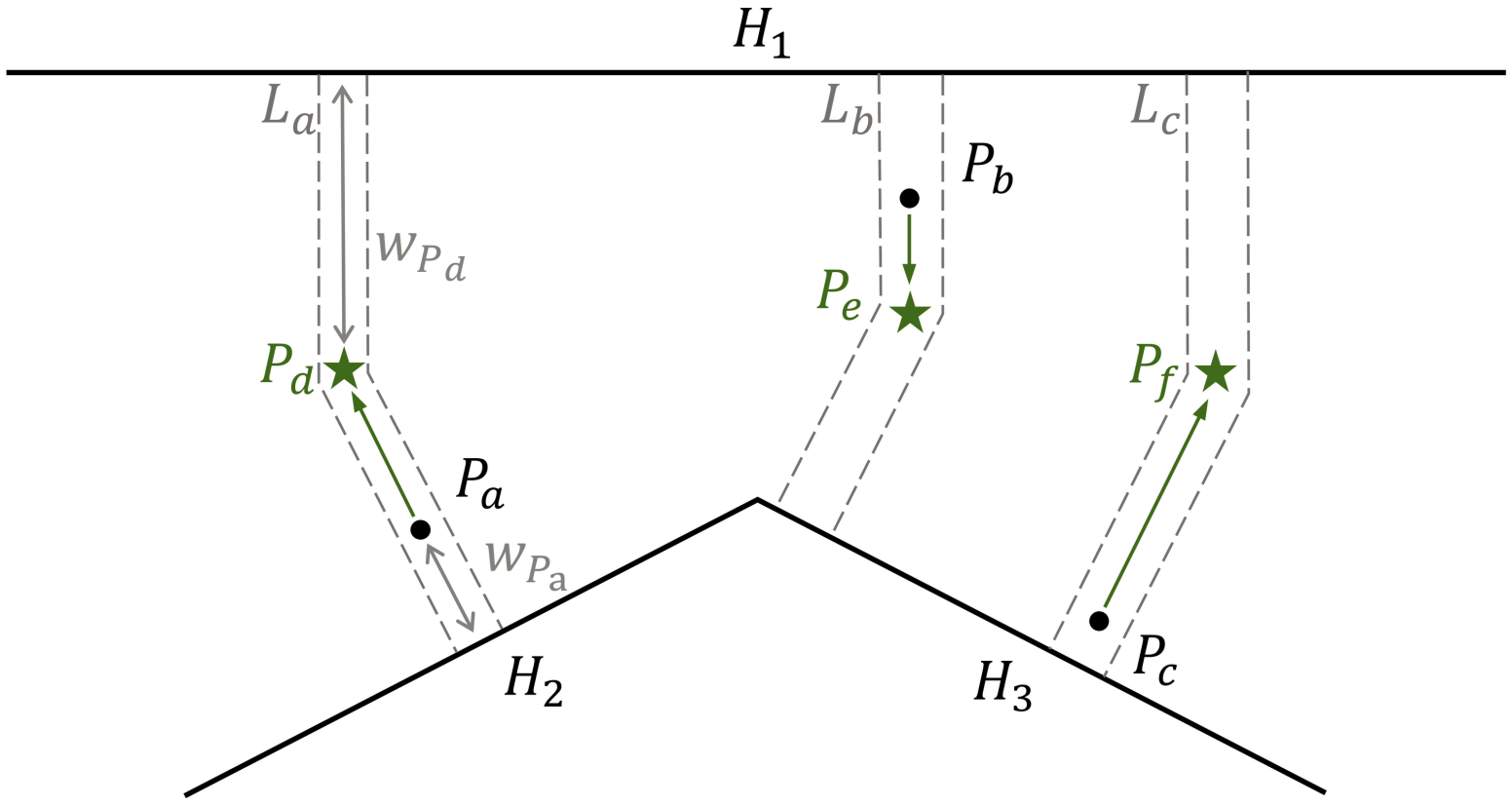

Figure 2. Illustrative Example of Dynamic Width Characteristic / The construction of cross-section $L_a$ and the corresponding dynamic width characteristics are illustrated by the movement of point $\mathrm{P}_a$ to $\mathrm{P}_d$.

Taking $L_a$ as an example, it is generated according to the local boundary geometries associated with $H_2$. By giving a point $\mathrm{P}_a$ as an initial state, the minimum distance from $\mathrm{P}_a$ to all geometric elements is denoted as $w_{P_a}$, which is associated with $H_2$. The point is then adjusted along the normal direction of $H_2$ until the distance to $H_2$ and $H_1$ become equal. The resulting point $P_a$ determines the dynamic width characteristic at this point, and the cross-section $L_a$ is generated. The detailed iterative computation of the dynamic width characteristic, including the determination of cross-sections $L$, is provided in in Section 2.2.

## 2.2. Methods for Calculating Dynamic Width Characteristic

### 2.2.1. Geometric Analysis for Boundary Elements

The method for calculating dynamic width characteristics optimizes the position of a virtual vessel relative to the geometric elements. The foundation of this method is first expressing how the distance changes as a virtual vessel moves relative to point or line features with geometric boundaries. A virtual vessel is introduced as a computational point whose position is updated in the plane with a prescribed step speed $v$ and direction $\phi(t)$, where $\phi \in [0, 2\pi)$. The position of the virtual vessel $P(t) = \left[x(t), y(t)\right]^T$ is governed by:

$$\frac{\mathrm{d}P(t)}{\mathrm{d}t} = v \cdot \left[\cos(\phi(t)), \sin(\phi(t))\right]^T, \tag{3}$$

where $t = 0$ is set as the initial time and $goal = \left[x_{end}, y_{end}\right]^T$ is the goal. The rates of change in distance between the virtual vessel and the line or point features are denoted as $\mathrm{d}D_{H_1} / \mathrm{d}t$ and $\mathrm{d}D_{H_2} / \mathrm{d}t$, see Figure 3 with a fixed line feature $H_1$ in Figure 3(a) and a fixed-point feature $H_2$ in Figure 3(b). Within an infinitesimal time interval $\mathrm{d}t$, the virtual vessel undergoes a small displacement $\mathrm{d}P$. As the virtual vessel moves along different direction angles $\phi$, the variation in the distance between the virtual vessel and the geometric elements, denoted as $\Delta D_{H_i}$, changes accordingly.

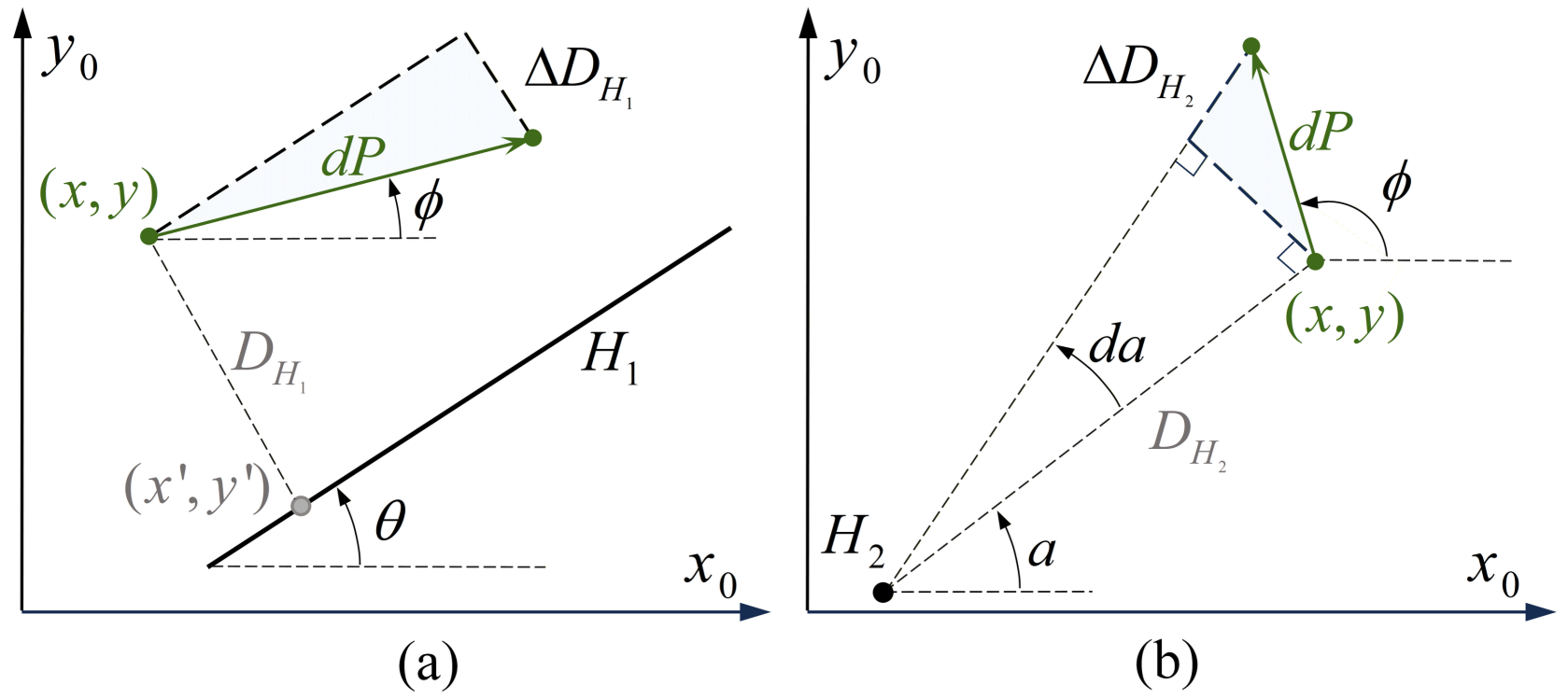

Figure 3. Illustration of the Relationship between a virtual vessel Movement and Geometric Elements/ (a) Line features; (b) Point features

For the line features, one has $\sin(\theta-\phi)=\mathrm{d}D_{H_1}/\mathrm{d}P$, where $\mathrm{d}P=v\cdot\mathrm{dt}$. Accordingly, the rate of change in the orthogonal distance $\mathrm{d}D_{H_1}/\mathrm{d}t$ between the virtual vessel and $H_1$ is given by:

$$\frac{\mathrm{d}D_{H_1}}{\mathrm{d}t}=\begin{cases}\operatorname{sign}\left((x-x')\cdot\csc(\theta)\right)\cdot\sin(\theta-\phi)\cdot v, \theta\neq 0\\ \operatorname{sign}(y-y')\cdot\sin(\phi)\cdot v, \theta=0\end{cases}, \tag{4}$$

where $\left[x', y'\right]^T$ represents the projection coordinates of the virtual vessel on $H_1$, and $\theta$ denotes the inclination angle of $H_1$ with $\theta\in[0,\pi]$.

For the point feature $H_2$, a similar approach yields:

$$\frac{\mathrm{d}D_{H_2}}{\mathrm{d}t}=\cos(\phi-\alpha)\cdot v, \tag{5}$$

where $\alpha$ represents the azimuth angle of the vector from $H_2$ to the virtual vessel, with $\alpha\in[0,2\pi)$. After the virtual vessel moves along the direction angle $\phi$ by a time step $\mathrm{d}t$, the infinitesimal change in $\alpha$ is given by:

$$\mathrm{d}\alpha=\frac{\sin(\phi-\alpha)\cdot v}{D_{H_2}}\mathrm{d}t. \tag{6}$$

Based on (4)-(5), the relationship between the iterative update of the virtual vessel and its distance to the geometric elements is established to calculate dynamic width characteristics.

### 2.2.2. Dynamic Width Characterization Function

Considering that sub-waterway units are categorized as 'line-line', 'point-line', or 'point-point' arrangements, we can calculate the dynamic width characteristics accordingly. Specially, suppose the boundaries of a sub-waterway unit can be decomposed into geometric elements $H_1$ and $H_2$. Without loss of generality, assume the ship's initial position is closer to $H_1$. Let the ship move at maximum speed away from $H_1$, using Eq. (4) or Eq. (5) to calculate the direction angle $\phi$ such that $\mathrm{d}D_{H_1} / \mathrm{d}t = v$. Continue the process until the minimum distances to the geometric elements are maximized, thereby obtaining the dynamic width characteristic as defined in **Definition 1**.

Subsequently, the virtual vessel continues moving to obtain the dynamic width characteristic at adjacent cross-sections within the sub-waterway unit, calculating the direction angle of the object's movement at each moment, satisfying

$$\frac{\mathrm{d}D_{H_1}}{\mathrm{d}t} = \frac{\mathrm{d}D_{H_2}}{\mathrm{d}t}. \tag{7}$$

This allows for the calculation of the dynamic variation in width at different positions within the sub-waterway unit, ultimately yielding the dynamic width characterization function. The dynamic width characterization function can be formalized as:

$$\omega_{dwc} = W(s), \tag{8}$$

where $s$ represents the arc length of the ship trajectory. For the ship moving at a velocity $v$, the arc length at any time $t$ is given by $s(t) = v \cdot t$. The variable $\omega_{dwc}$ is the dynamic width characteristic at that point on the trajectory corresponding to the arc length $s$.

By establishing the dynamic width characterization function for each sub-waterway unit and then combining them, the dynamic width characterization function for the entire waterway can be determined by Algorithm 1.

**Algorithm 1** Dynamic Width Characterization Function for Sub-waterway Units

**Define:** geometric elements $H_i (i \in \{1,2\})$, ship location $P(t) = [x(t), y(t)]^{\mathrm{T}}$ and movement direction $\phi(t)$;
**Initialize:** $t = 0$, $\Delta t = 1\mathrm{s}$, $v = 1\mathrm{m/s}$, $goal = [x_{end}, y_{end}]^{\mathrm{T}}$, an empty list $R$;
**while** $dist(P(t), goal) > 0$ **do**
  Calculate $\frac{\mathrm{d}D_{H_i}}{\mathrm{d}t}$ according to (4)-(5);
  Calculate $w_{P(t)} = \min(D_{H_1}, D_{H_2})$;
  **if** $D_{H_1} < D_{H_2}$ **then**
    Calculate $\phi(t)$ such that $\frac{\mathrm{d}D_{H_1}}{\mathrm{d}t} = v$;
  **else if** $D_{H_1} > D_{H_2}$ **then**
    Calculate $\phi(t)$ such that $\frac{\mathrm{d}D_{H_2}}{\mathrm{d}t} = v$;
  **else**
    Calculate $\phi(t)$ such that $\frac{\mathrm{d}D_{H_1}}{\mathrm{d}t} = \frac{\mathrm{d}D_{H_2}}{\mathrm{d}t}$;
    Append $w_{P(t)}$ to the list $R$;
  **end if**
  Update $P(t+\Delta t) \leftarrow P(t) + v \cdot \Delta t \cdot \begin{bmatrix} \cos(\phi(t)) \\ \sin(\phi(t)) \end{bmatrix}$;
  $t \leftarrow t + \Delta t$;
**end while**
**Return** the dynamic width characterization function based on $R$;

### 2.2.3. Waterway Intersection

The intersection of an intricate waterway presents a special scenario when calculating dynamic width characteristics, as it involves more than two geometric boundaries. Consider the case of three intersecting waterways, see Figure 4. By following the previous process, we identify a point $P_a$ at the intersection. This point is equidistant from the upper, lower, and right geometric boundaries, representing the largest minimum distance to these geometric boundaries. Using $P_a$ as a reference, the waterway can be divided into three distinct sections.

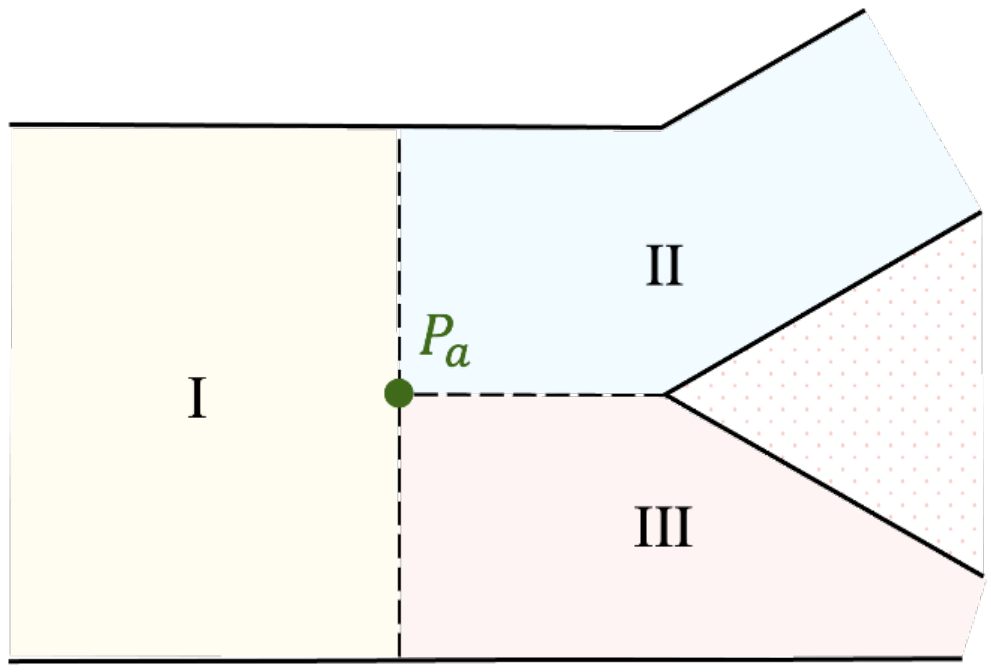

Figure 4. Sketch of the waterway intersection condition.

More specifically, as the virtual object moves through Waterway I, the nearest point and line features associated with each sampled position $P(t)$ are identified. Let $N(P(t))$ denotes the number of nearest geometric features at $P(t)$. When $N(P(t)) > 2$, the current position is identified as a waterway-intersection point and denoted as $P_a$. Upon reaching $P_a$, the ship can enter either Waterway II or Waterway III. If it chooses Waterway II, the dynamic width characterization function is computed based on the upper and right boundaries. Similarly, the function for Waterway III can be determined. This is applicable to the intersections in both geographical contexts and those caused by traffic or other barriers (e.g., shallow waters, or buoys). By tracking the ship's movement and identifying points where it has more than two nearest geometric elements, it is possible to detect intersections and compute the dynamic width characterization function for intersecting waterways effectively.

## 3. Geometric Analysis-based Safety Assessment

This section proposes the Geometric Analysis-based Route Safety Assessment (GARSA) framework, see Figure 5, to be applied in narrow, rugged, and intricate waterway environments. GARSA explores all feasible routes and establishes the corresponding dynamic width characterization functions. To facilitate the selection of the safest route, a safety performance index is introduced, which converts the dynamic width characterization functions of each route into numerical scores for easy comparison.

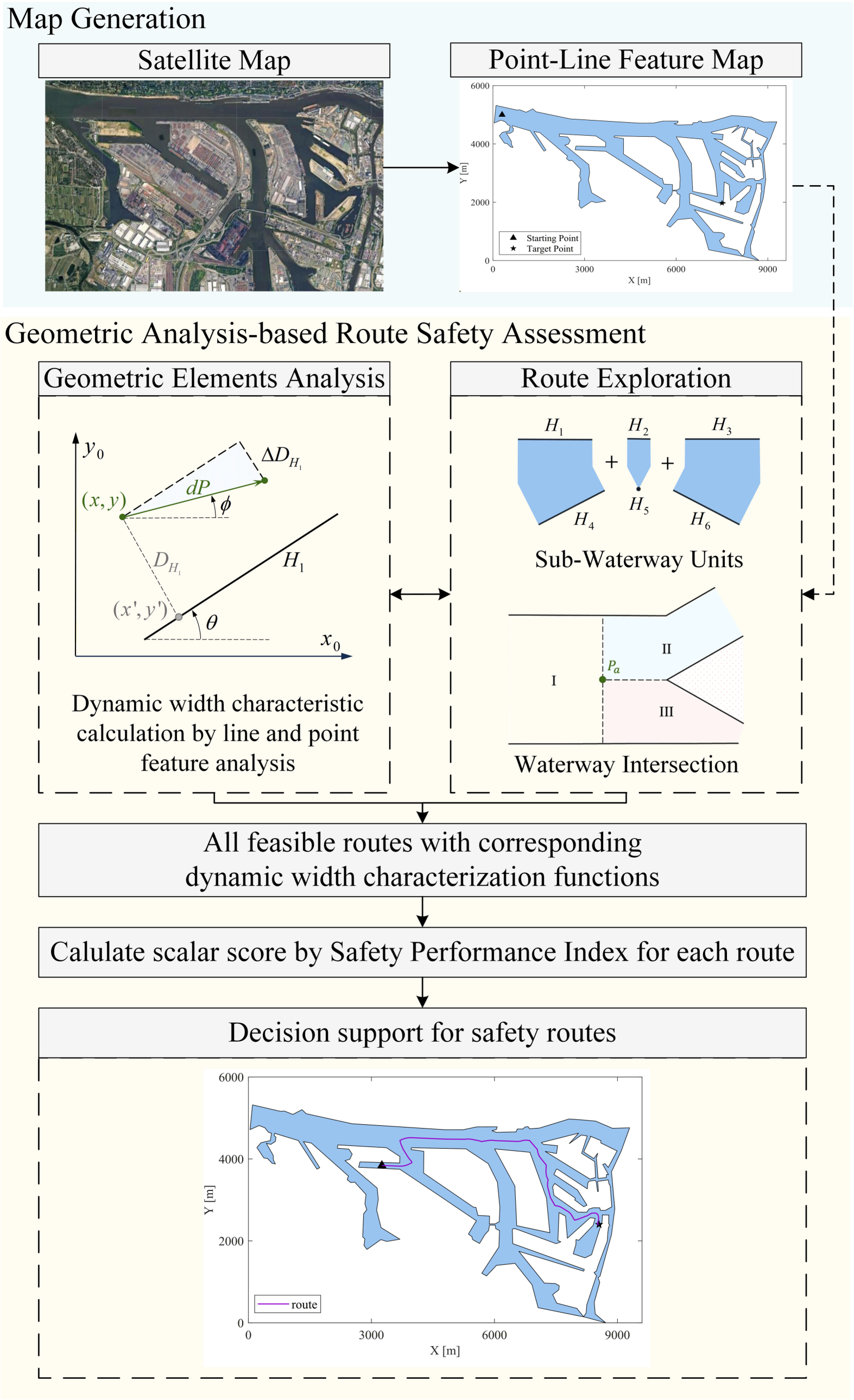

Figure 5. Diagram of the Safety Route Decision Support Process.

### 3.1. Dynamic Width Characterization Function Calculation

GARSA calculates the dynamic width characteristic by utilizing point and line boundary features, see Figure 6. This calculation can be seen as a process of object movement, where intersections and connectivity relationships of sub-waterway units are identified, as discussed in Section 2. In this process, intersections are represented as nodes, and every branching pathway must be navigated. A pathway is fully explored once the object either reaches the terminus of the waterway or encounters a node that has already been explored. By traversing all sub-waterway units and intersections, the exploration of all possible routes can be conducted while establishing the corresponding dynamic width characterization function.

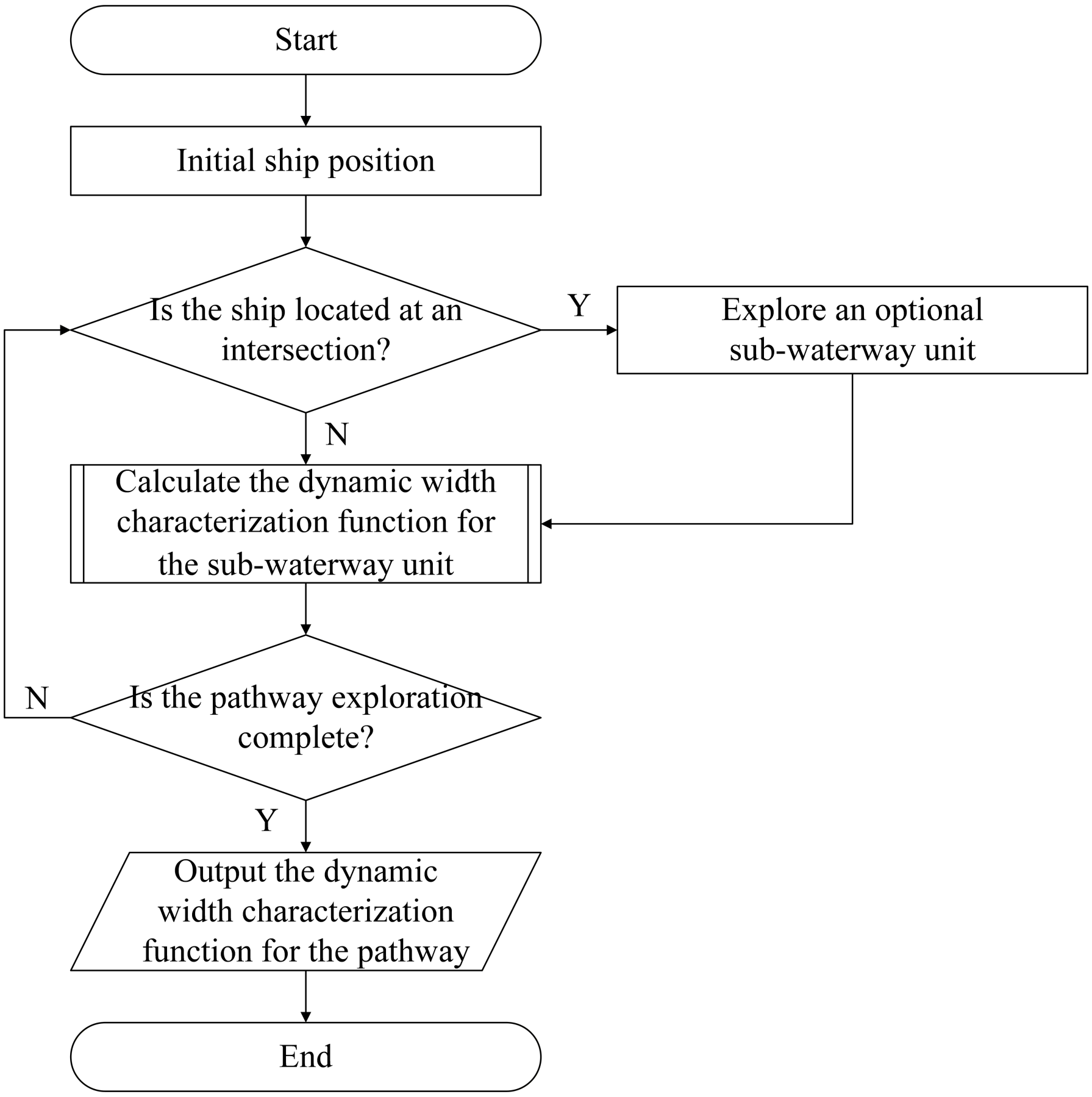

Fig. 6. Dynamic Width Characterization Function Calculation

### 3.2. Safety Performance Index

Navigational safety can be assessed and reflected in a decision support system by a safety performance index (SPI) that converts the dynamic width characterization functions into numerical scores. The SPI is formulated as a ratio, where the numerator represents the overall navigable space, while the denominator characterizes the risk induced by locally narrow channel segments. This formulation enables the SPI to simultaneously capture navigational safety at both global and local levels: when the channel is globally wide, the SPI attains a higher value, indicating favorable navigational safety. In contrast, the presence of narrow segments along the route increases the denominator, thereby reducing the SPI and highlighting these localized high-risk areas. Suppose the trajectory generated by the object is discretized into $n$ points. The SPI is defined as:

$$\kappa_{sp} = \frac{\bar{\omega}_{dwc}}{\sum_{i=1}^{n} -\log_{10}\left(\frac{W(s_i)}{\xi}\right) \cdot \mathrm{I}_{\{W(s_i)<\xi\}} + \varepsilon}, \tag{9}$$

where $\bar{\omega}_{dwc}$ represents the average of dynamic width characteristics along the waterway. The logarithmic term in the denominator $\bar{\omega}_{dwc}$ nonlinearly amplifies the risk contribution of narrow sections. The indicator function $\mathrm{I}$ identifies narrow sections along waterways, such that a segment is classified as narrow when the dynamic width characteristics $W(s_i)$ is smaller than the threshold $\xi$ that is a defined parameter related to the required safety of vessel navigation and is mainly influenced by vessel size and maneuverability.

A small regularization term $\varepsilon$ is added to the denominator as a basis to avoid a zero denominator when there is no section along the route. This term is set to a small positive value with limited effect on the SPI ranking under normal conditions. In summary, when SPI is applied to evaluate a route, the dynamic width characterization function is obtained from the procedure illustrated in Figure 6.

This ratio-based formulation inherently balances global navigational safety and local risk. A route with generally wide waterways but with some highly constrained sections will yield a lower index, highlighting the local risks even in globally safe waters. This property makes the index particularly useful for comparing different routes. For example, a route with a consistent, moderate width throughout may have a higher SPI than a route that features wide sections interspersed with narrow, risky areas. By providing a quantifiable measure of navigational safety, this index aids in decision-making processes. It can guide autonomous ships in selecting safer routes, considering both overall route characteristics and localized risks. Additionally, it can also serve as a monitoring tool for navigational authorities, helping them implement safety measures in areas with lower SPI values due to narrow and intricate waterway sections.

## 4. Validation

### 4.1. Route Network Modeling by Geometric Analysis

To validate the effectiveness of the proposed method, the Port of Hamburg and another hypothetical restricted waterway are utilized as the test environment. The Port of Hamburg is characterized by numerous intersections, intricate connectivity between sub-waterway units, and multiple narrow, rugged passages. The remote sensing map of the waterways (see Fig. 7(a)) is pre-processed into point-line feature maps (see Fig. 7(b)). On this basis, the route decision-making based on the proposed GARSA framework is tested and analyzed comprehensively.

Firstly, the route network modeling function is illustrated in subsection 4.1. Secondly, we explore the safety assessment capabilities for three scenarios in subsection 4.2. These scenarios include the port entry and in-port navigation processes in the Port of Hamburg, as well as navigation in a hypothetical restricted waterway.

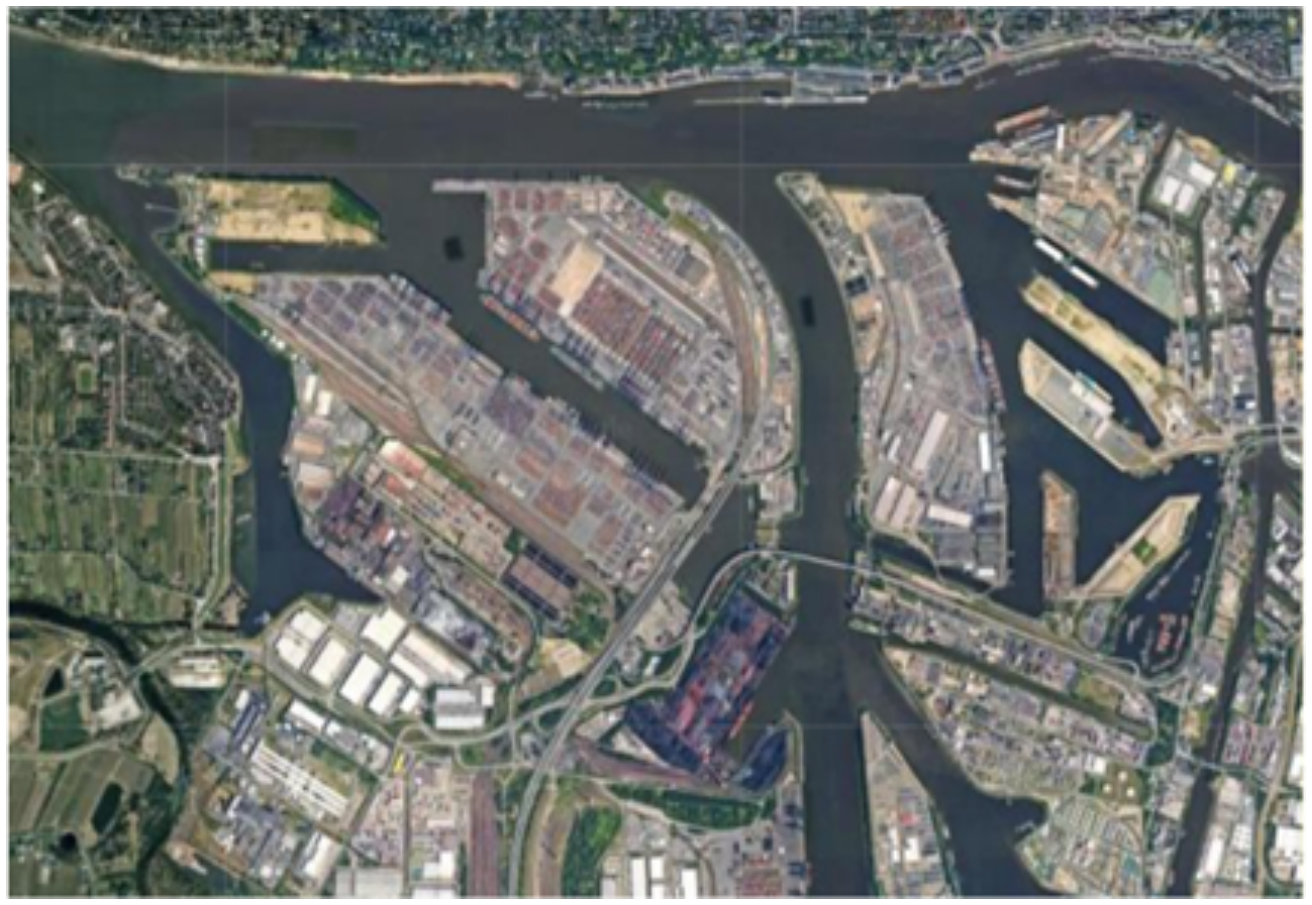

(a) Map of Hamburg Port

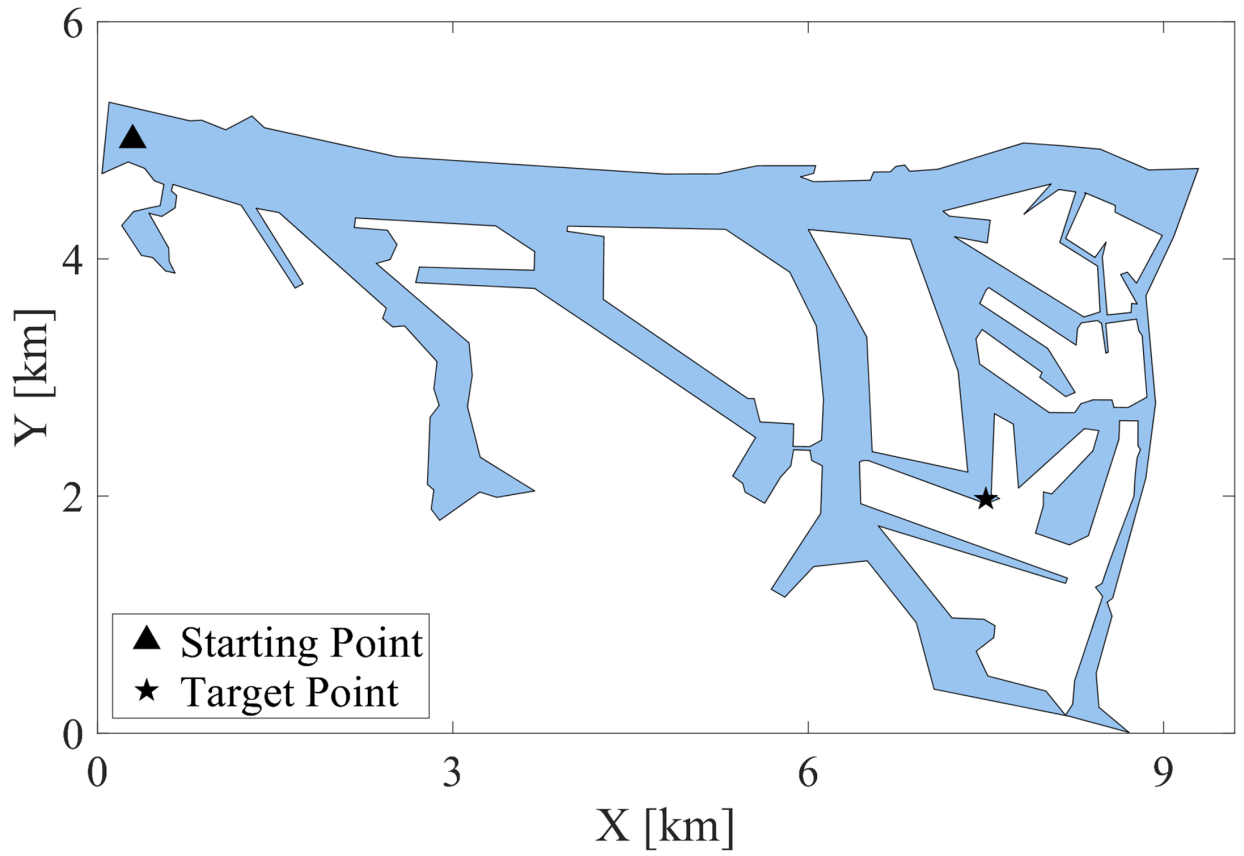

(b) Point-Line Feature Map

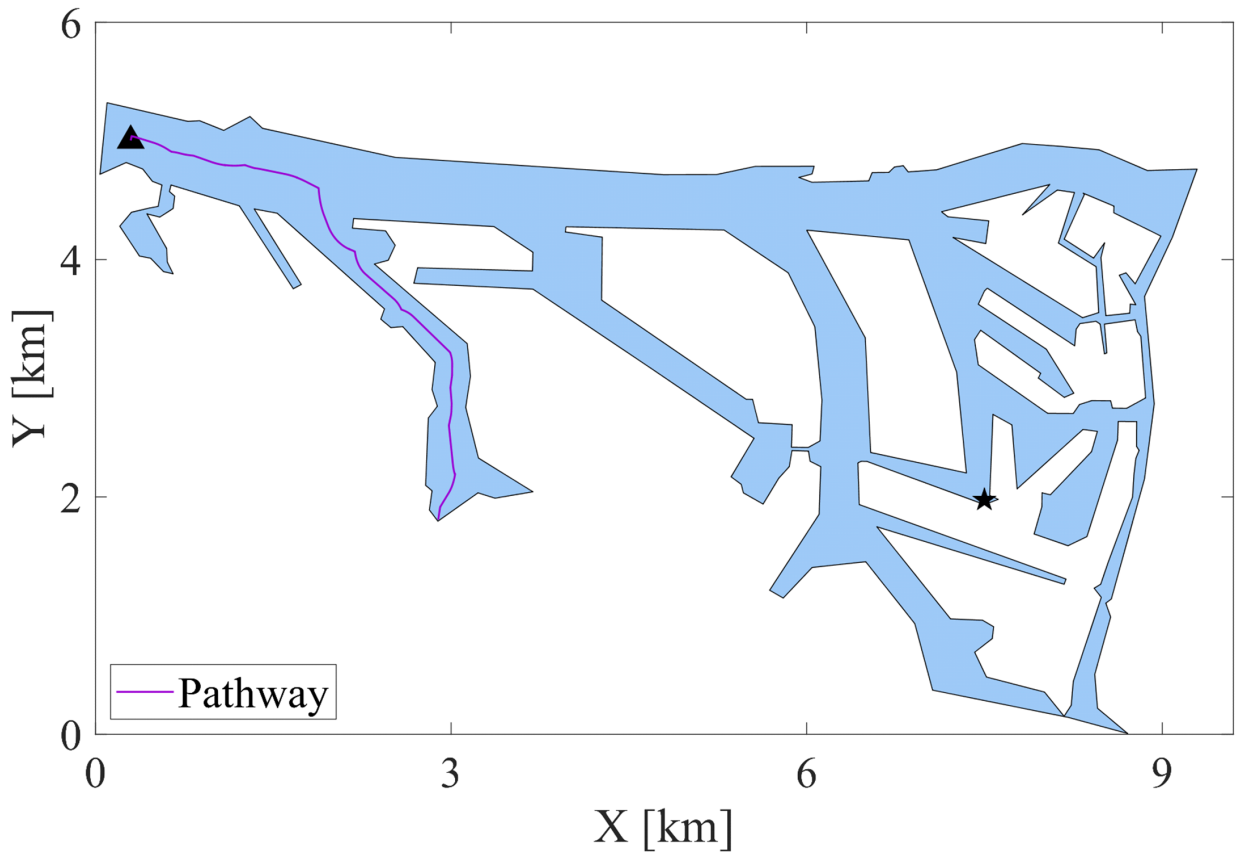

(c) Sample Pathway 1

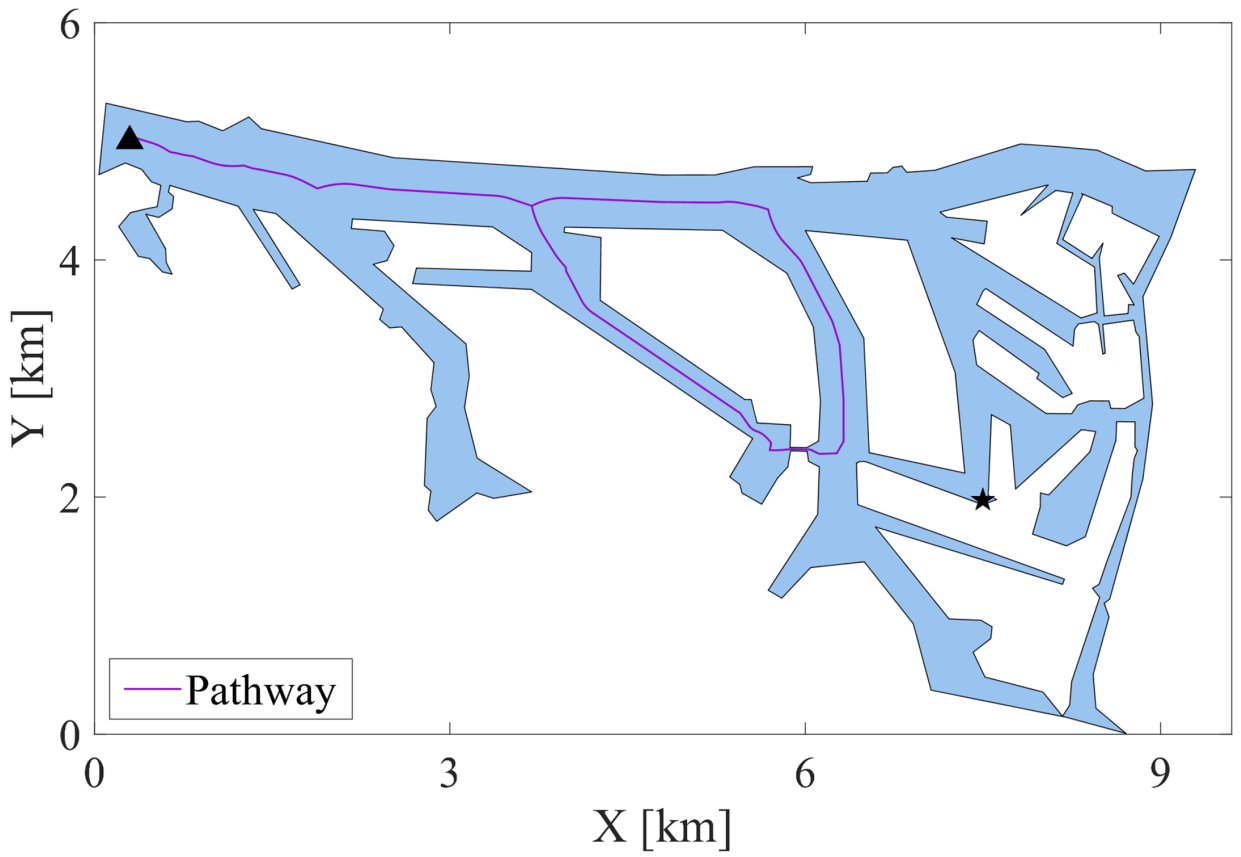

(d) Sample Pathway 2

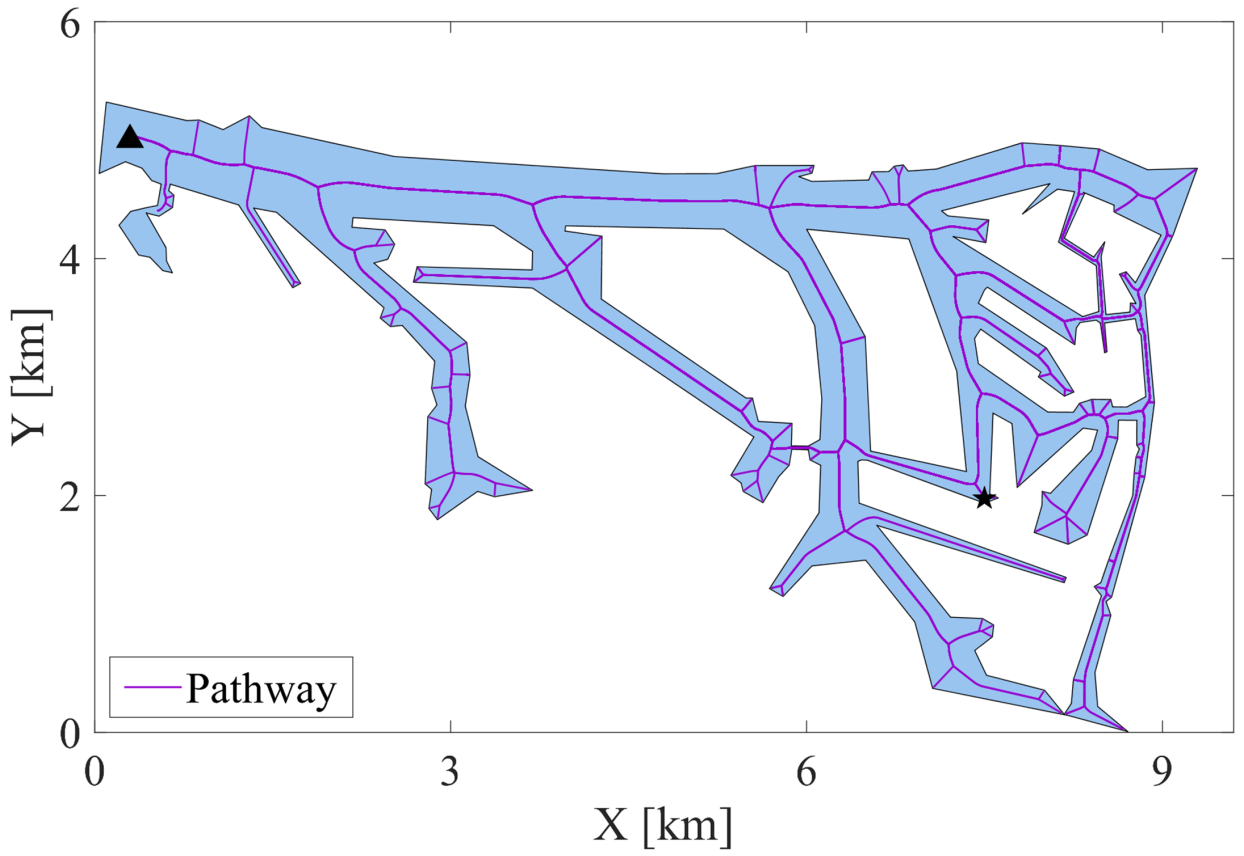

(e) All Pathways

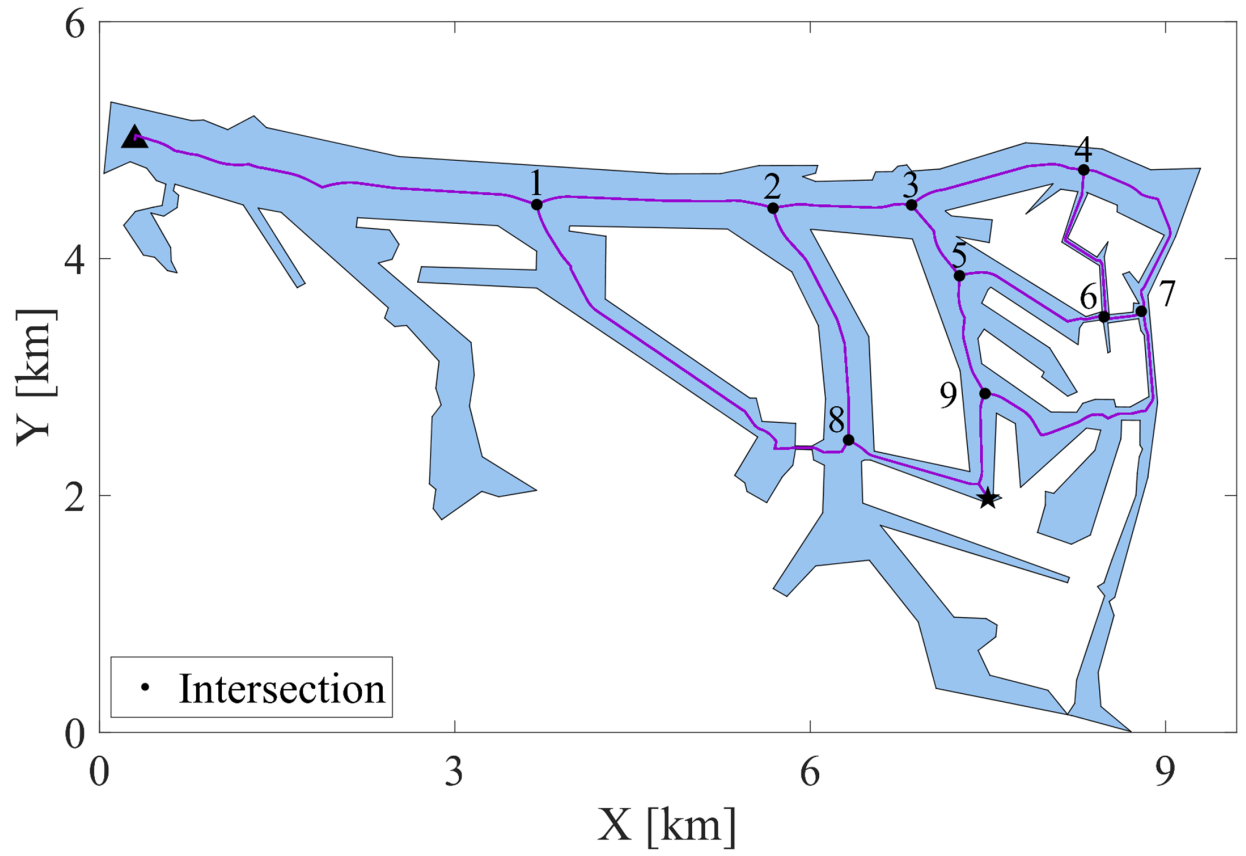

(f) Routes to Target Point

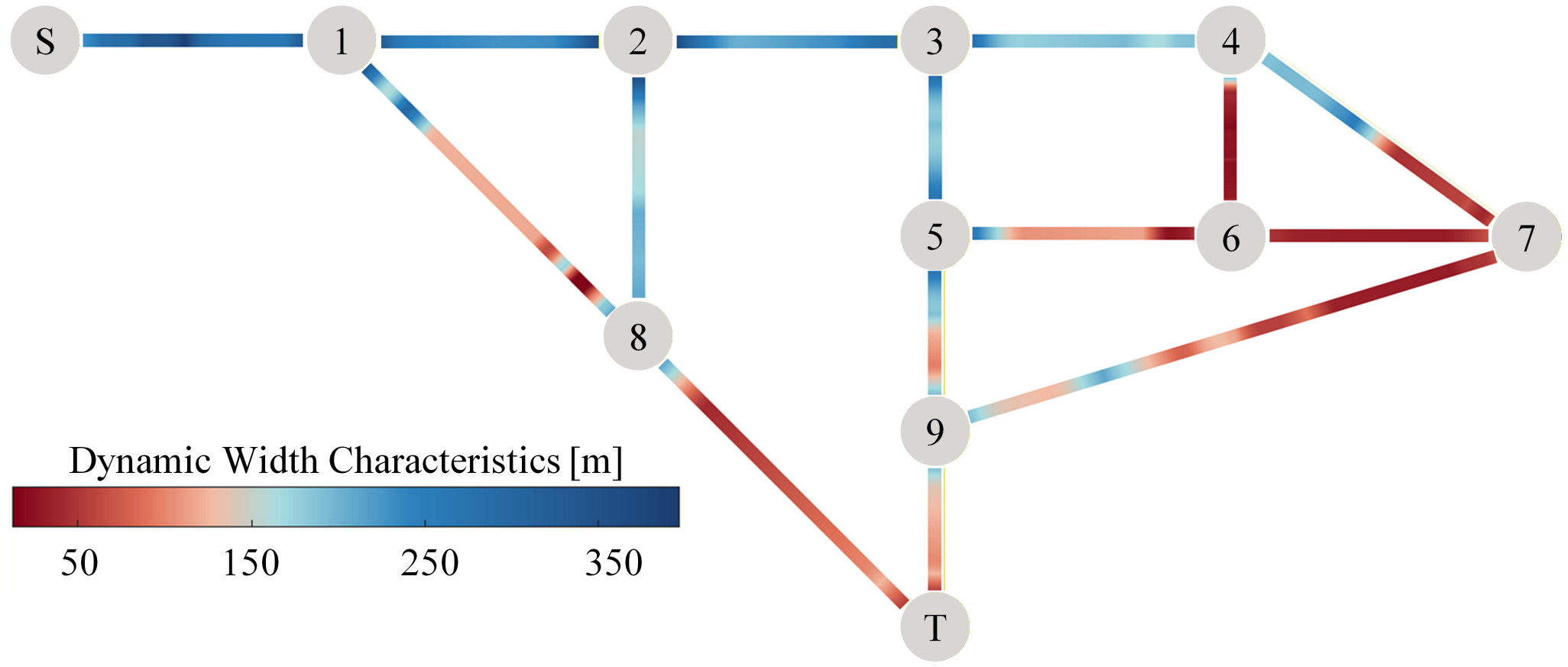

(g) Simplified Diagram of Intersections

Figure 7. Snapshots of Route Exploration in the Port of Hamburg.

Figure 7 shows the route exploration within the Port of Hamburg conducted by the GARSA framework. The exploration is carried out through the iterative calculations of object's movement, as described in Section 2. This process provides an initial representation of all potential pathways within the waterway, while simultaneously capturing dynamic width characteristics. It is assumed that the waterway boundaries ensure that the water depth at any point exceeds the potential ship's draft, making the entire area navigable.

The exploration begins with an initial position, with a ship moving along the waterways while keeping as far as possible from nearby boundary geometric elements. In this process, intersections are treated as nodes, and all branches need to be traversed. A complete pathway is considered explored when the object either reaches the end of the waterway or encounters a previously explored node. Figures 7(c) and 7(d) illustrate two sample pathways. Once all pathways have been explored, as shown in Fig. 7(e), one can filter the possible routes to a target point by identifying the pathway that connects the starting point to the target point. In this example, a total of 16 feasible routes is identified, traversing eight intersections within the waterway system, see Fig. 7(f). This is the initial map representation process from the perspective of geometric analysis.

Figure 7(g) presents the route network model, revealing the connectivity and spatial properties between the intersection nodes. The dynamic width characteristics are illustrated by a heat map along the connection pathways, where the color gradient from blue to red reflects the varying widths. Blue indicates wider sections, suggesting areas with potentially lower navigational risks and greater ease of transit. In contrast, darker red represents narrower, more constrained spaces that may pose higher navigational challenges or risks, thereby quantifying the variation in the width of the waterways. Consistent with the map visualization, the pathways between nodes 6 and 7, nodes 4 and 6, and from node 8 to the target exhibit narrower areas that require careful consideration. The extracted connectivity and width data provide valuable insights into the spatial dynamics of the waterway, supporting further risk assessment and decision-making in navigation planning.

### 4.2. Decision Support for Safety Waterways

In this subsection, we investigate the safety-oriented decision-support capability of GARSA. Unlike existing studies [17], [28] which typically simplify waterways into interconnected node networks to improve vessel scheduling and navigation efficiency, this study analyzes the navigational risks encountered by ships in irregular waterways from a localized and scenario-specific perspective. The representative scenarios considered include port entry and intra-port navigation in the Port of Hamburg, as well as navigation within a hypothetical restricted waterway.

The overall procedure follows the workflow illustrated in Fig. 5, which consists of route exploration, computation of dynamic width characterization functions, and safety assessment using the Safety Performance Index (SPI), with the threshold set to $\xi = 70$. In this subsection, navigation zones recommended by maritime conventions and regulations are not considered; instead, the geometric characteristics of the waterways are used as the test environment to evaluate the performance of the proposed method. This assumption is reasonable, as the point-line feature map can be further refined in future work by incorporating existing navigation rules and zonal constraints.

#### 4.2.1. Case 1: Port entry

Figures 8 (a)-(e) present five explored routes as illustrative examples. Intuitively, Routes 1 and 2 pass through the upper right region of the map, where narrower waterways increase the risk of collisions. This is a result of GARSA exploring all possible pathways. In contrast, Routes 3, 4, and 5 are depicted with better navigational safety, although Routes 3 and 4 also pass through several constricted areas.

Figure 8(f) presents the dynamic width characterization functions for Routes 3, 4, and 5, providing detailed insights into the relationship between route length and the varying widths along each route. Specifically, Route 3 encounters multiple narrow and potentially hazardous areas, with dynamic width measurements at the route lengths of 6,200, 6,600, and 7,300 meters being 64, 14, and 40 meters, respectively. In contrast, Route 4 exhibits a dynamic width of 40m at the 7800-meter mark, which aligns with the 7300-meter point on Route 3. Comparatively, Route 5 maintains a dynamic width greater than 70m throughout, ensuring a consistently safer navigational path up to the target point.

The navigational safety of each route is quantified, as shown in Table 1. Route 5 emerges as the safest option with the highest SPI value. This finding aligns with the map visualization, which illustrates a favorable width profile and fewer narrow sections along the route. In contrast, Routes 1 and 2 exhibit the lowest SPI values, primarily because they traverse extended narrow areas. Although their average dynamic widths are acceptable, these constricted segments significantly degrade their overall safety performance. Route 3 also has a low SPI, despite having the shortest route length. This is attributed to its path through several extremely narrow areas, with a minimum dynamic width of only 14 meters. This illustrates that SPI is not influenced by route length. While Routes 4 and 5 have comparable average widths, Route 5's SPI is notably higher, reflecting the effectiveness of SPI in accounting for local narrow regions. This demonstrates the sensitivity of SPI as a tool for assessing navigational safety across different routes.

Table 1. Comparison of Navigational Safety Across Inbound Routes

| Route | Mean DWC (m) | Minimum DWC(m) | Route Length (m) | SPI |
|---|---|---|---|---|
| 1 | 188 | 18 | 13229 | 0.22 |
| 2 | 199 | 18 | 12865 | 0.33 |
| 3 | 194 | 14 | 8395 | 0.97 |
| 4 | 230 | 41 | 8848 | 4.07 |
| 5 | 240 | 55 | 9286 | 150.86 |

* DWC refers to the Dynamic Width Characteristic, SPI refers to Safety Performance Index; * $\xi$=70m.

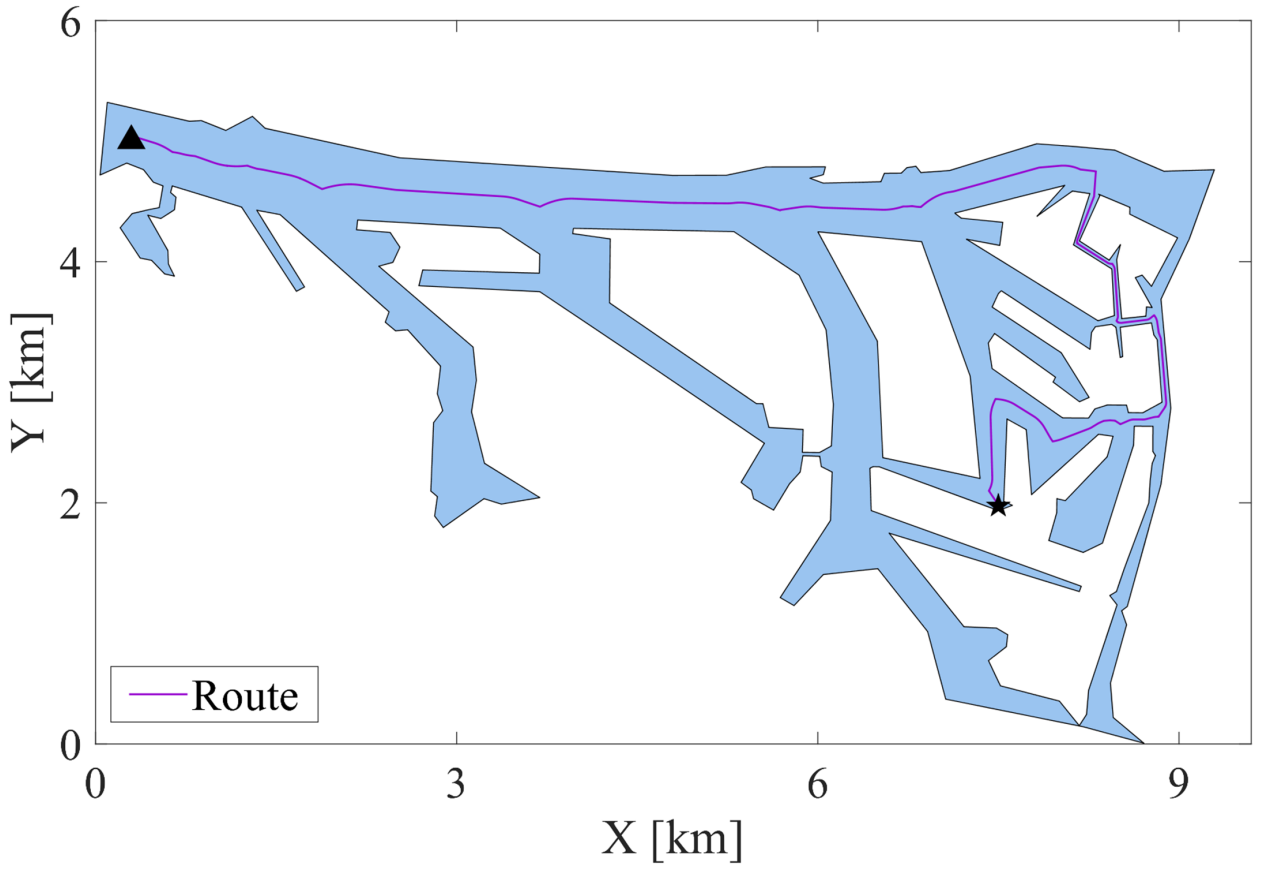

(a) Route 1

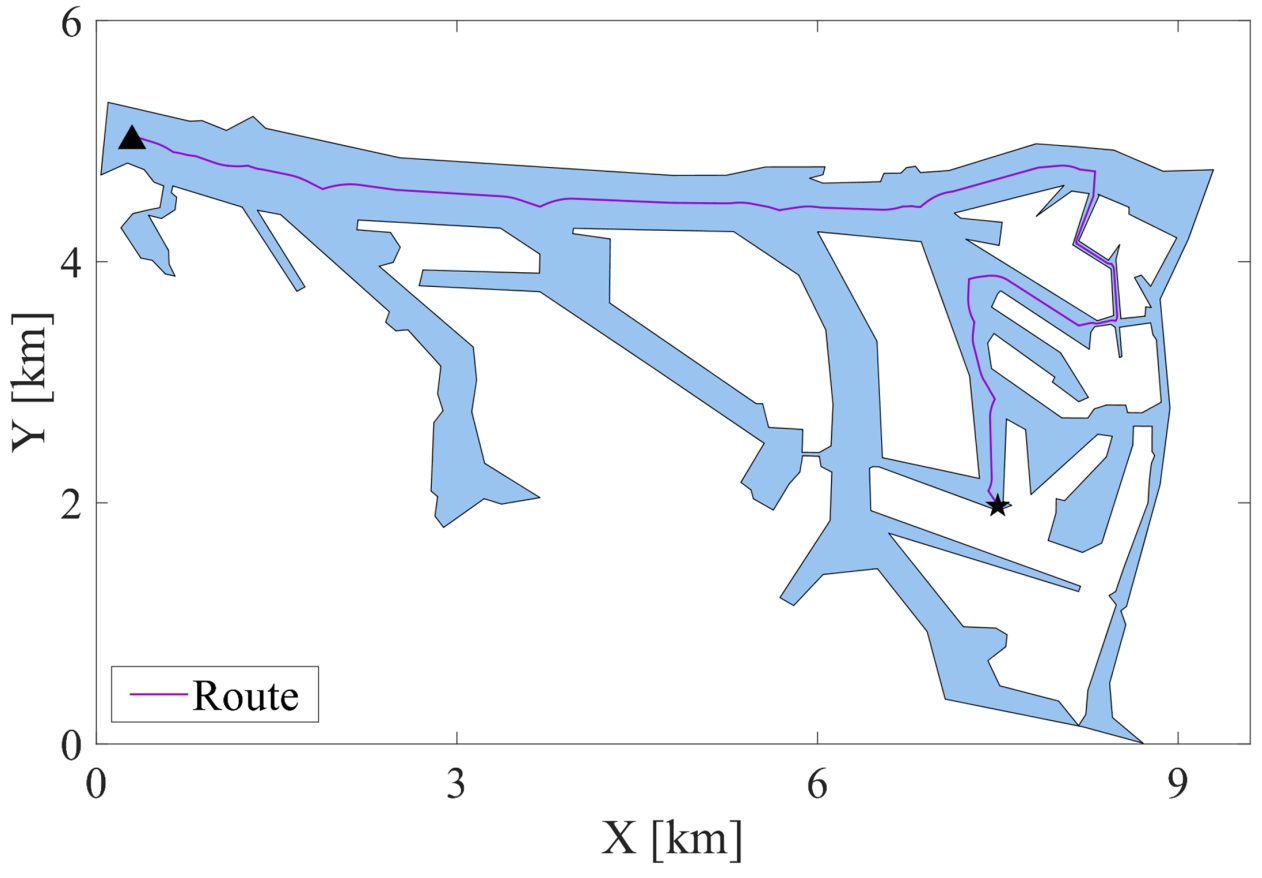

(b) Route 2

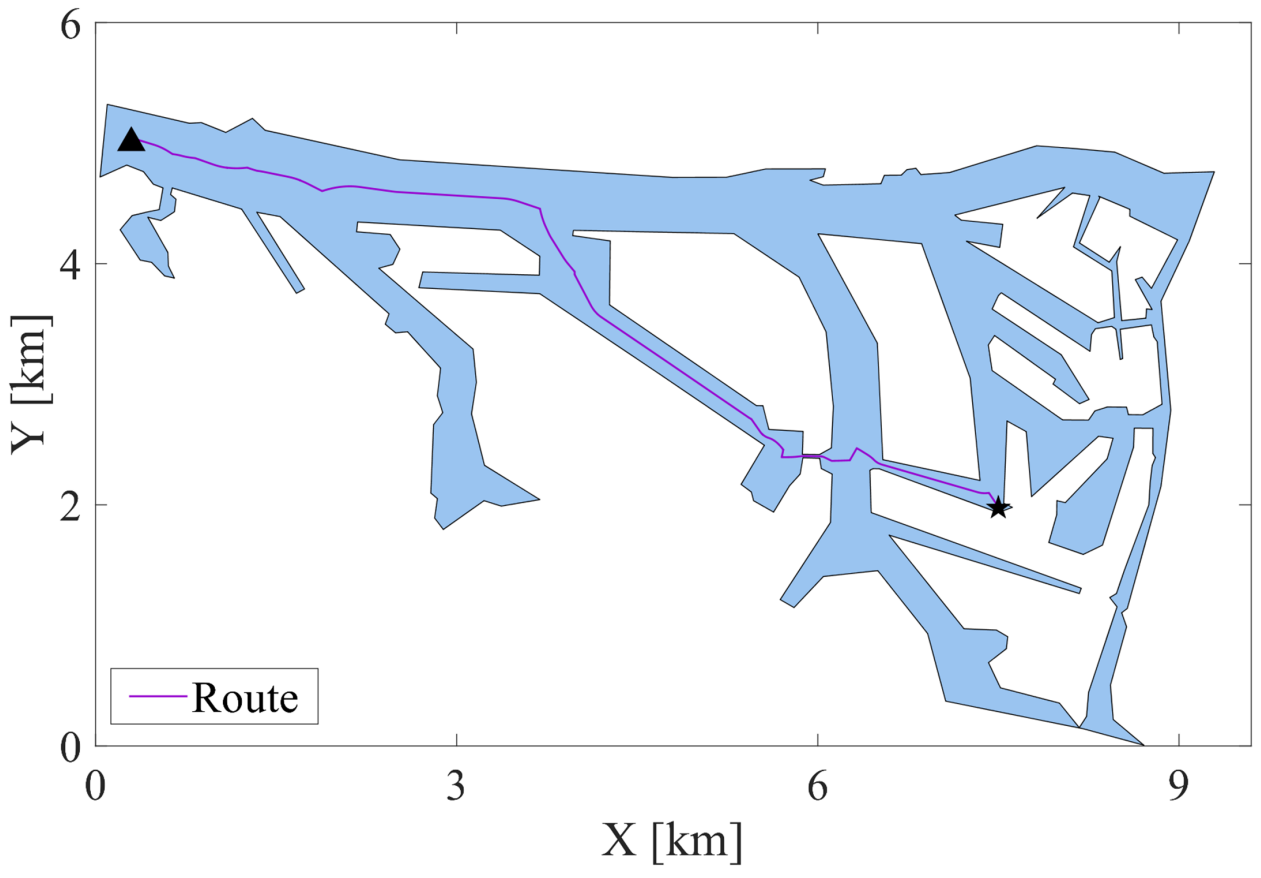

(c) Route 3

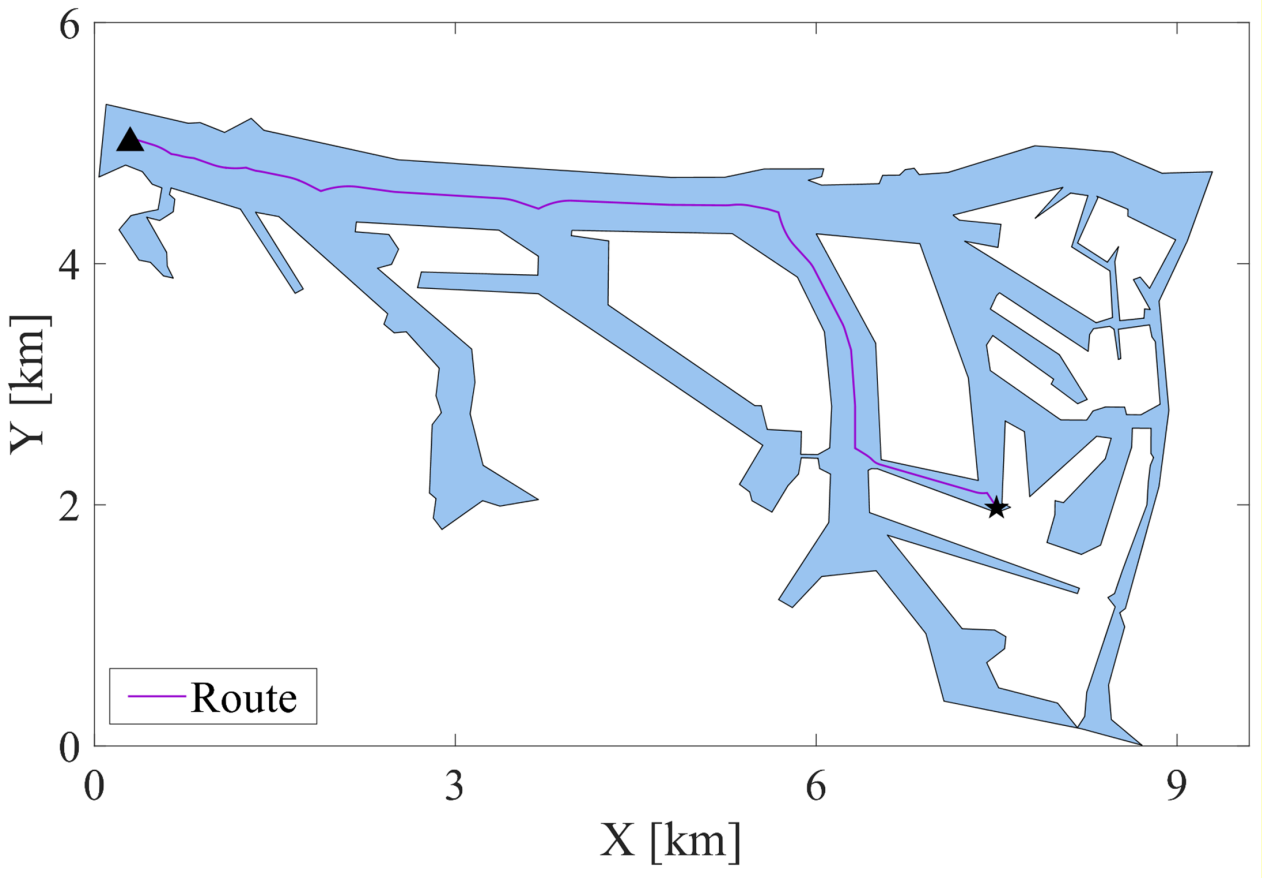

(d) Route 4

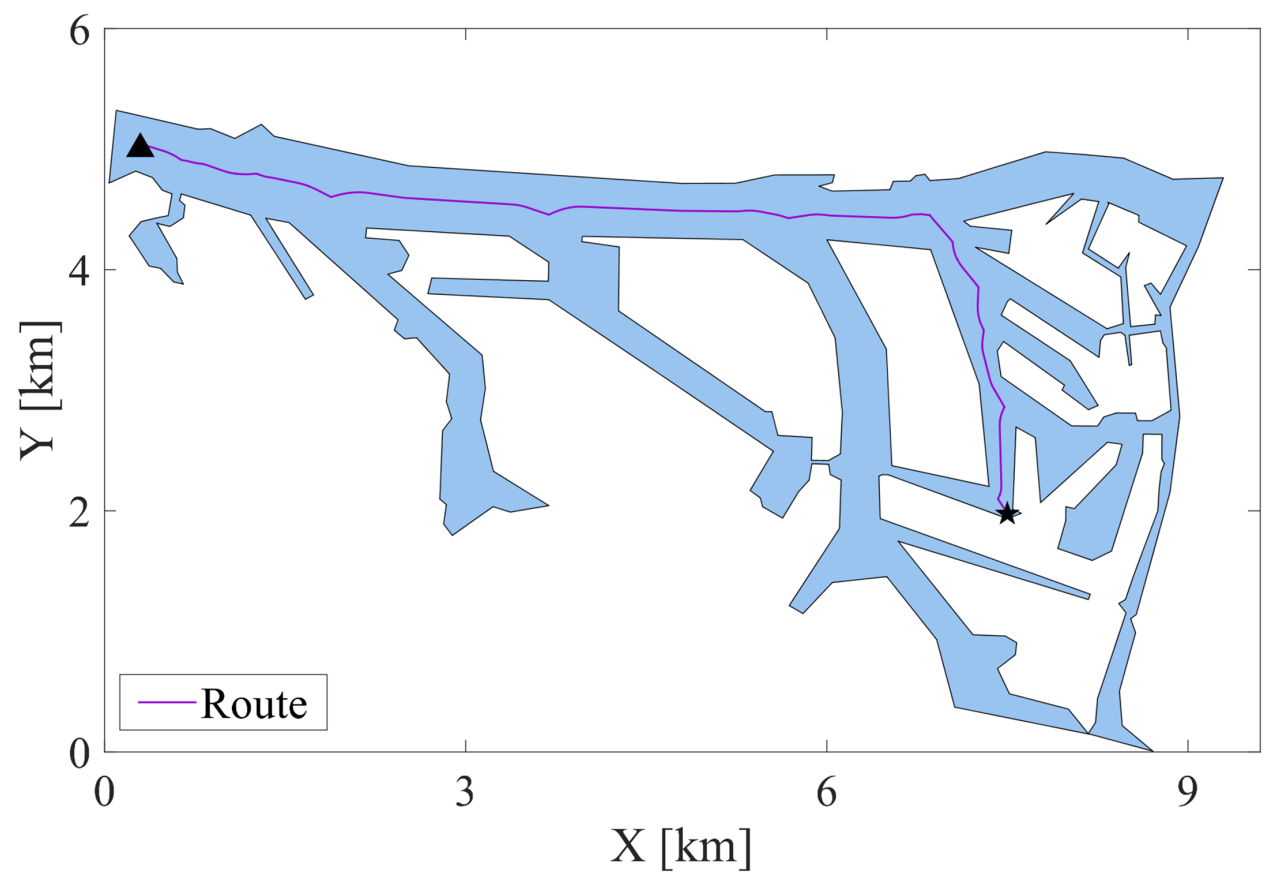

(e) Route 5

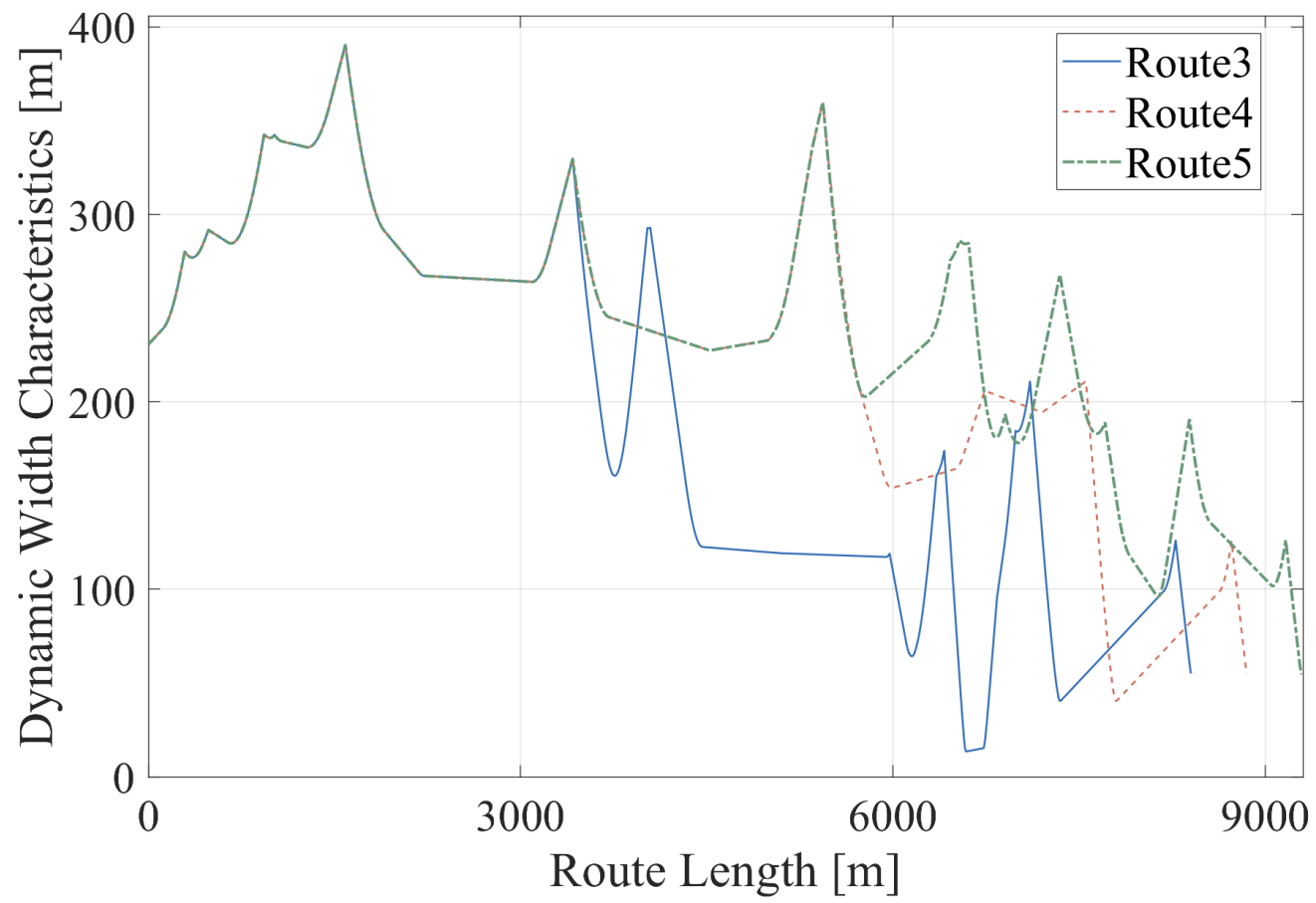

(f) Dynamic Width Characterization Function

Figure 8. Snapshots of Port Entry Route Exploration and Assessment.

#### 4.2.2. Case 2: Intra-port navigation

Given the starting and target points illustrated in Figure 9, the proposed GARSA framework explores 24 potential routes. Table 2 shows the safety assessment results using the SPI. Route 1, the narrowest with a mean dynamic width of 117 meters and a minimum dynamic width of 14 meters, has a low SPI of 0.59, indicating significant navigational challenges. Routes 3 and 4 are similar in mean dynamic width and minimum dynamic width, both are significantly better than Route 1, Route 4's SPI of 135.10 is substantially higher than Route 3's SPI of 3.06. This discrepancy suggests that Route 3 may have some localized narrow areas compared to Route 4, consistent with the results from map visualizations, implying a balance of global navigational safety and local risk. Therefore, through the analysis of the GARSA framework, Route 4 is identified as the safest route.

Overall, the results demonstrate the consistency of the GARSA framework in supporting safety route decision-making across various navigational tasks. The SPI offers a quantifiable and effective method for evaluating navigational safety, where higher scores are associated with wider routes and fewer narrow sections, providing intuitive insight into the relative safety of each route. By utilizing this approach, decision-makers can prioritize routes that minimize navigational risks, particularly in complex and constrained waterways like those found in the Port of Hamburg.

Table 2. Comparison of Navigational Safety Across Inter-Port Transit Routes

| Route | Mean DWC (m) | Minimum DWC (m) | Route Length (m) | SPI |
|---|---|---|---|---|
| 1 | 117 | 14 | 7258 | 0.59 |
| 2 | 161 | 14 | 10382 | 1.11 |
| 3 | 172 | 41 | 8956 | 3.06 |
| 4 | 195 | 50 | 7845 | 135.10 |

* DWC refers to the Dynamic Width Characteristic, SPI refers to Safety Performance Index; * $\xi = 70m$.

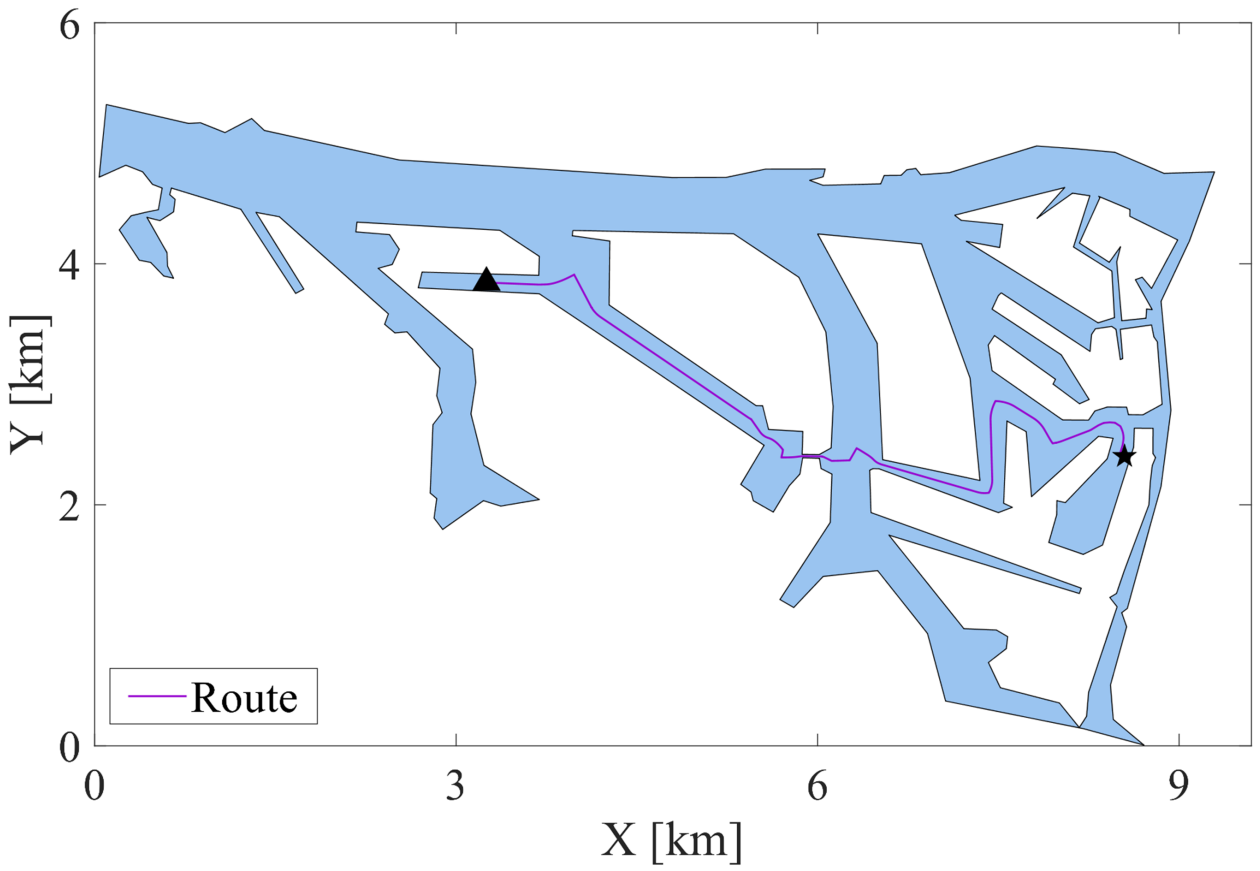

(a) Route 1

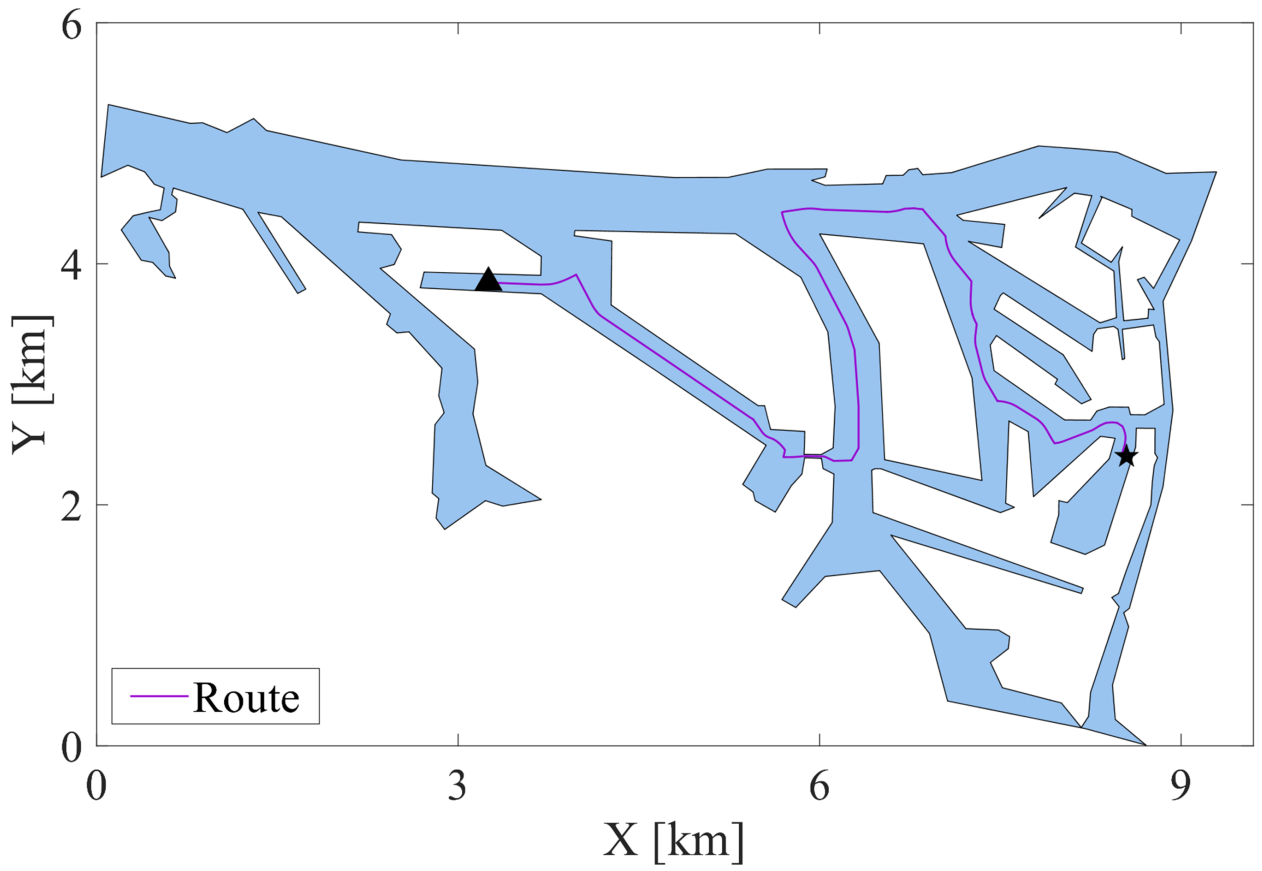

(b) Route 2

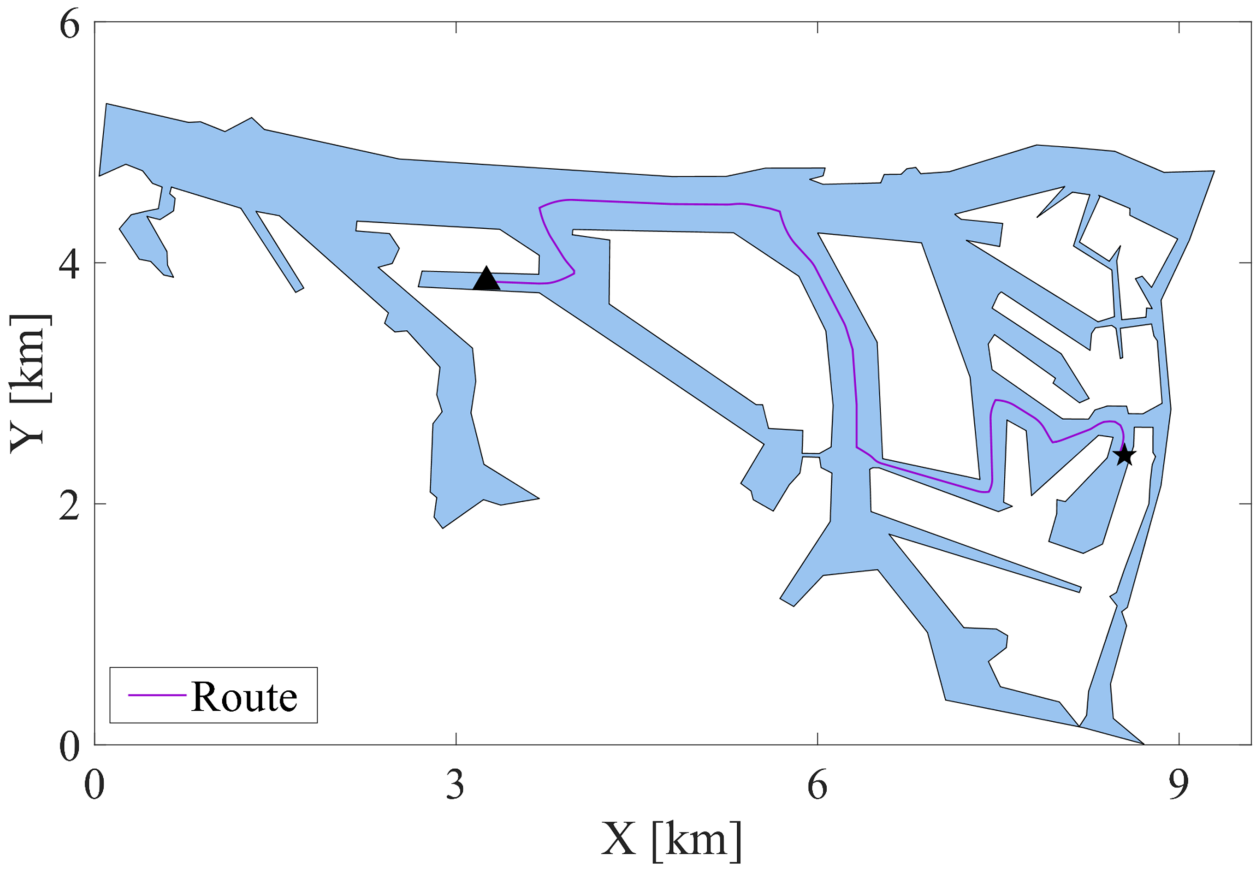

(c) Route 3

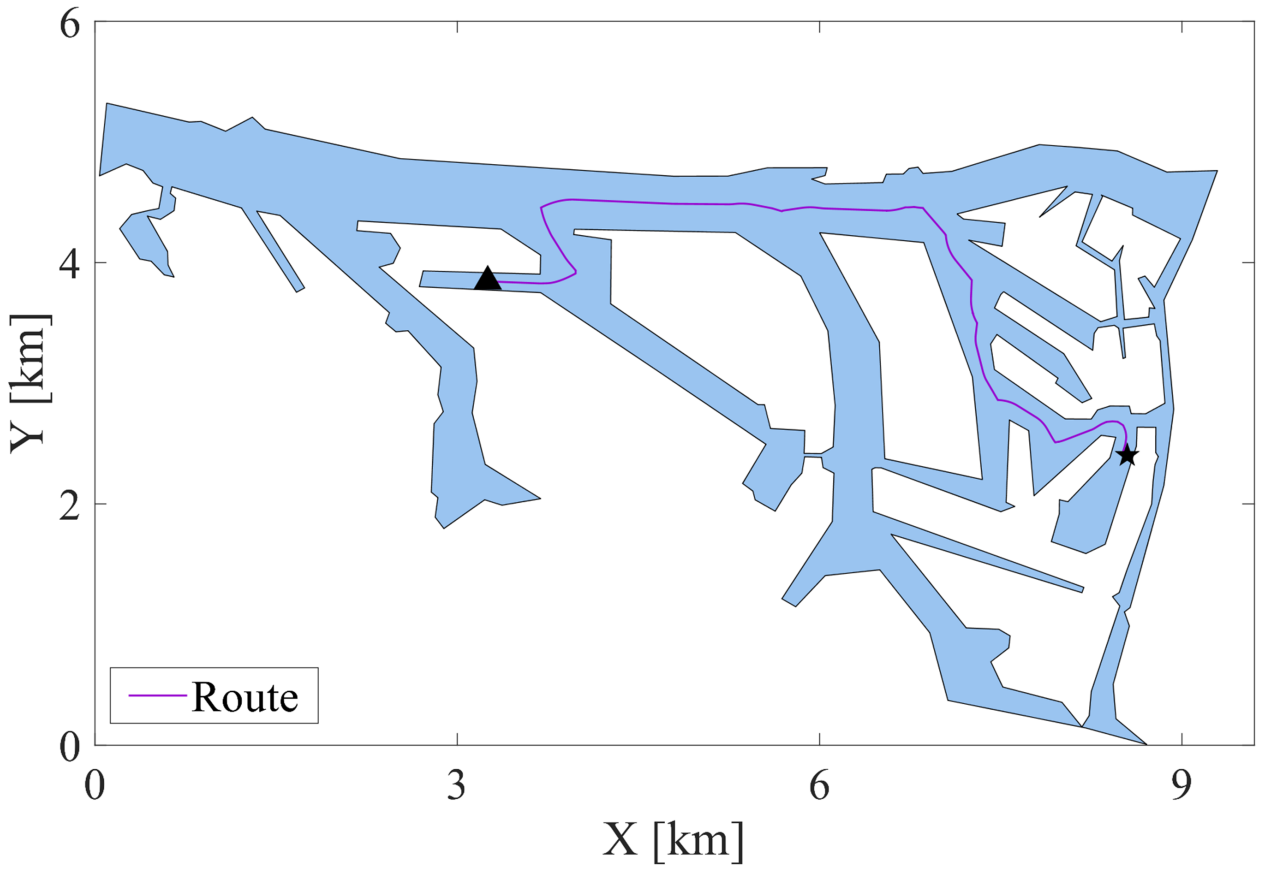

(d) Route 4

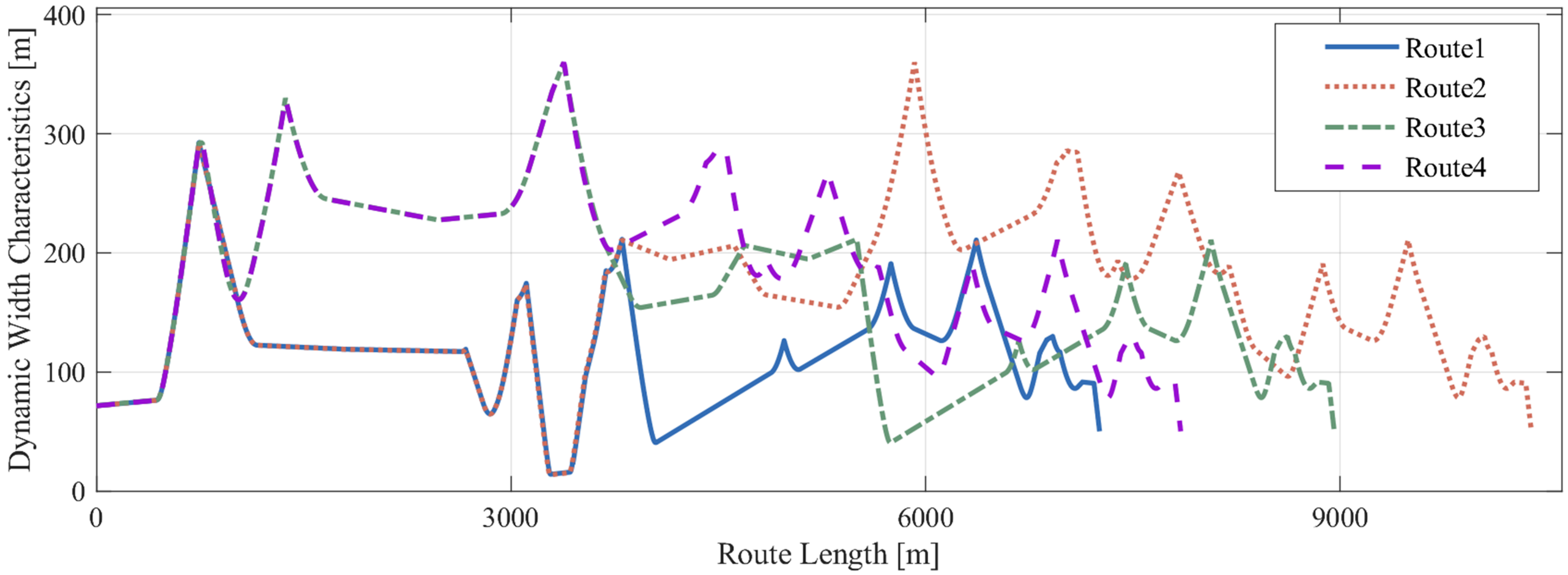

(e) Dynamic Width Characterization Function

Fig. 9. Snapshots of Inter-port Transit Route Exploration and Assessment

4.2.3. Case 3: Navigation in a hypothetical restricted waterway

To validate the adaptability of the GARSA framework in waterways with different geometric structures, Figure 10 presents three routes in a new scenario. Visually, compared with the geometric structure of the Hamburg Port waterways, these three routes have relatively similar overall navigable spaces, with no clearly identifiable extremely narrow sections. Specifically, Table 3 lists the SPI scores for the three routes. At a threshold of $\xi = 150$, the navigational safety of routes 1, 2, and 3 decreases sequentially. The SPI scores are 1.21, 0.62, and 0.61, respectively. Although route 1 has a slightly lower average dynamic width than routes 2 and 3, the decrease in SPI for routes 2 and 3 is mainly due to their passage through a relatively narrow area. The SPI is constructed as a ratio that highlights the risk of narrow sections. This design allows Route 1 to maintain relatively high scores, despite its slightly lower average dynamic width.

Table 3. Comparison of Navigational Safety in a hypothetically restricted waterway

| Route | Mean DWC (m) | Minimum DWC(m) | Route Length (m) | SPI |
|---|---|---|---|---|
| 1 | 198 | 103 | 11638 | 1.21 |
| 2 | 221 | 81 | 13041 | 0.62 |
| 3 | 220 | 81 | 15536 | 0.61 |

* DWC refers to the Dynamic Width Characteristic, SPI refers to Safety Performance Index.
* $\xi = 150m$.

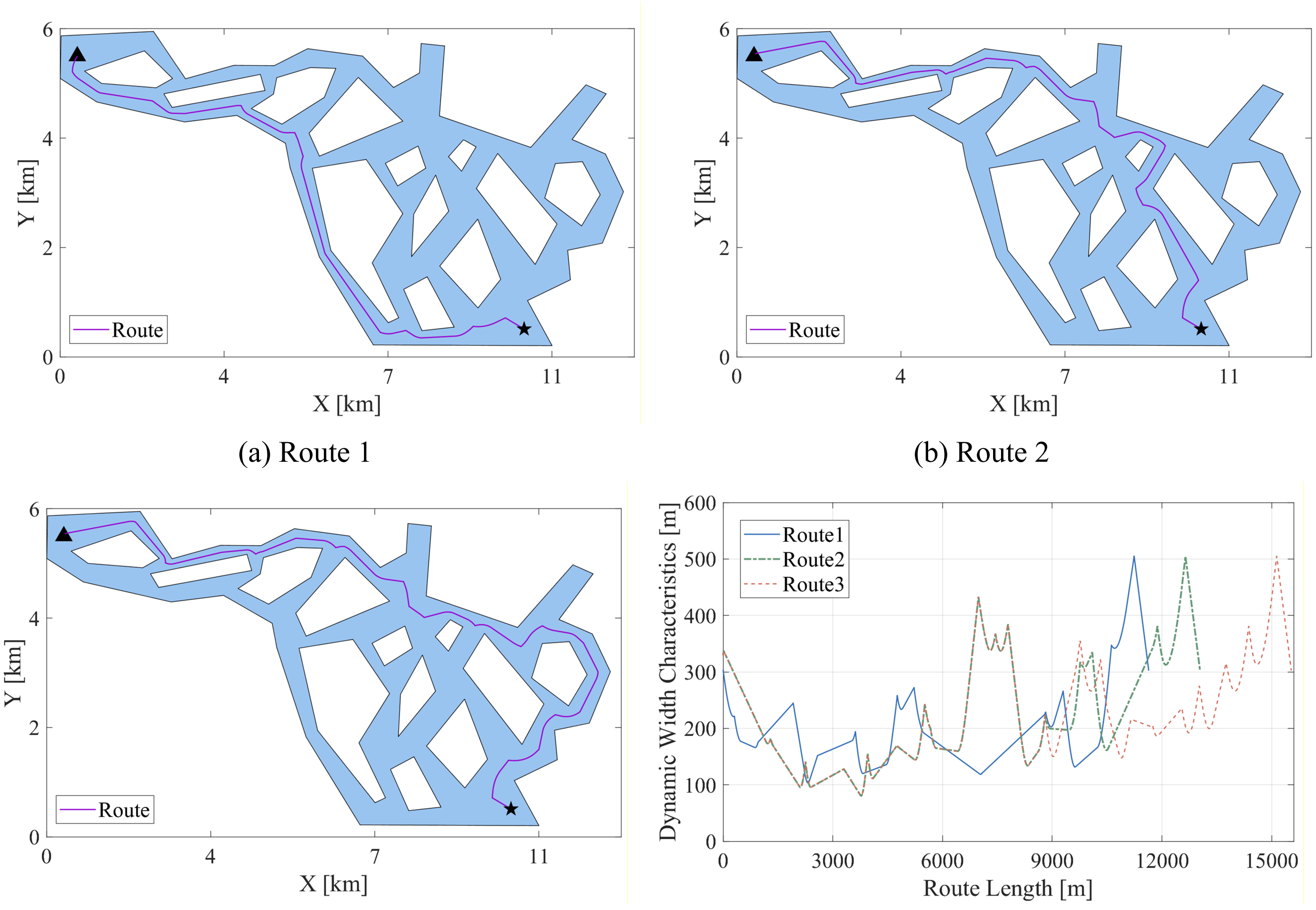

(a) Route 1

(b) Route 2

(c) Route 3

(d) Dynamic Width Characterization Function

Figure 10. Snapshots of navigation under a hypothetical restricted waterway

### 4.3. Sensitivity Analysis of the Threshold ξ on SPI

To analyze the impact of the threshold $\xi$ on the SPI, SPIs are computed under different threshold settings that identify narrow sections along a route and are related to the safety domain required for navigation, which is mainly influenced by vessel size and maneuverability. Table 4 presents the SPIs of the Hamburg Port inbound routes, the inter-port transit routes, and the hypothetical scenario under different threshold values. The SPIs decrease with an increased $\xi$. However, the route with the highest SPI remains unchanged in all three scenarios: Route 5 for the Hamburg Port inbound routes, Route 4 for the inter-port transit routes, and Route 1 for the hypothetical scenario. Figure 11 further illustrates the variations in SPI under different threshold settings. This indicates that the main route selection results are robust to the tested range of $\xi$.

The sensitivity of SPI differs among routes because of their different dynamic width characteristic distributions. Routes containing a larger proportion of constrained sections tend to exhibit lower sensitivity of SPI to variations in the threshold $\xi$, since these sections are consistently identified as narrow across the tested threshold range. Overall, although the absolute SPIs are affected by the choice of $\xi$, the main route ranking and decisions remain consistent within the tested threshold ranges.

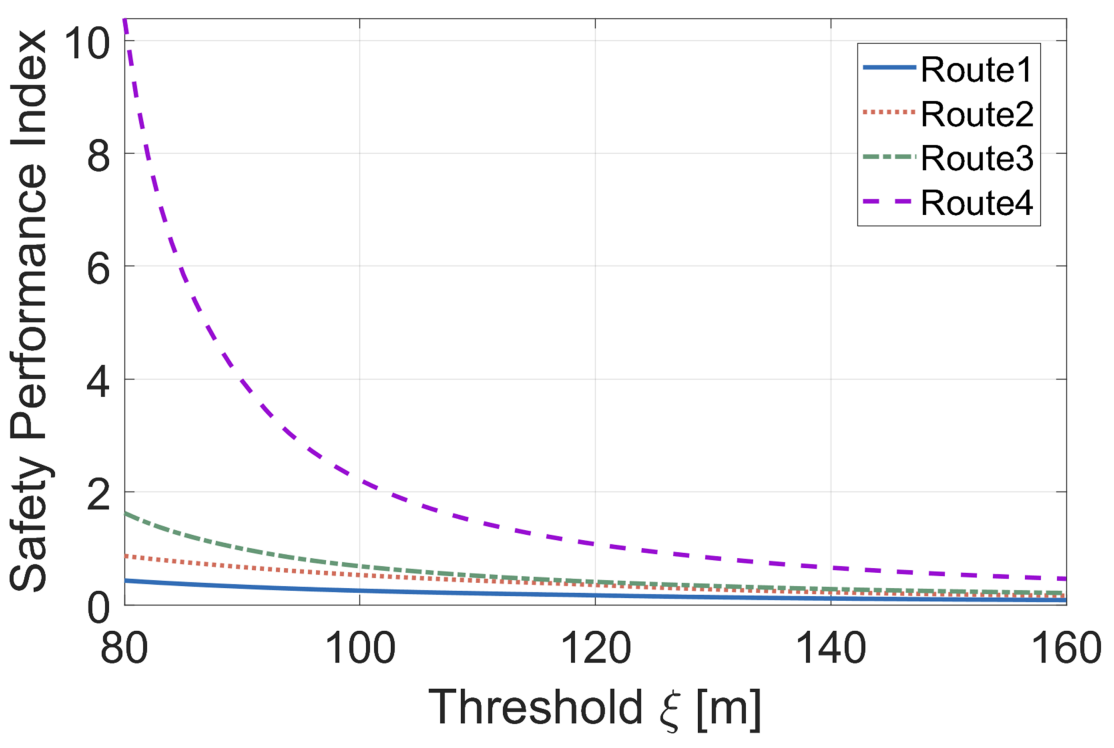

(a) Inter-Port Transit Routes

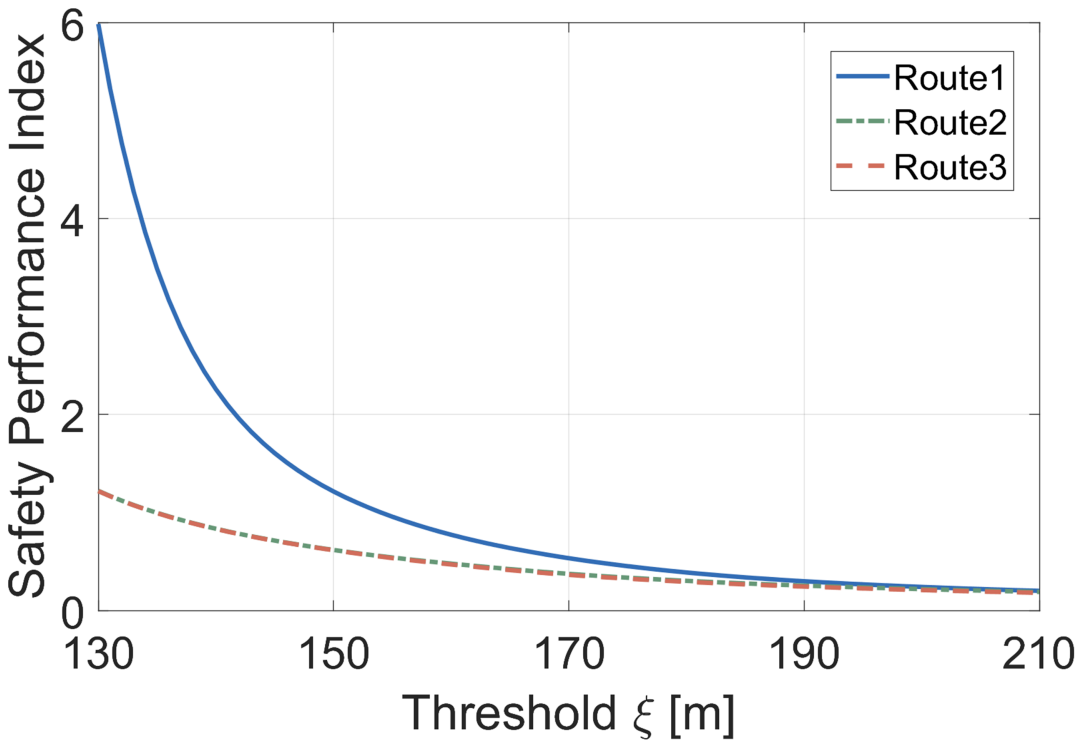

(b) Hypothetical Scenario Routes

Figure 11. Illustration of SPI Variations at Different Thresholds

Table 4. SPI Values at Different Thresholds

| (a) SPI Values for Inbound Routes at Different Thresholds | | | | | |
|---|---|---|---|---|---|
| $\xi$ | Route 1 | Route 2 | Route 3 | Route 4 | Route 5 |
| 60 | 0.27 | 0.40 | 1.28 | 8.22 | **985.68** |
| 80 | 0.18 | 0.28 | 0.75 | 2.52 | **63.88** |
| 100 | 0.14 | 0.23 | 0.49 | 1.32 | **22.69** |
| 120 | 0.12 | 0.19 | 0.36 | 0.88 | **4.99** |
| 140 | 0.10 | 0.15 | 0.25 | 0.67 | **2.05** |
| 160 | 0.09 | 0.13 | 0.20 | 0.55 | **1.25** |
| (b) SPI Values for Inter-Port Transit Routes at Different Thresholds | | | | | |
| 60 | 0.78 | 1.30 | 6.16 | **466.27** | |
| 80 | 0.43 | 0.87 | 1.63 | **10.39** | |
| 100 | 0.25 | 0.53 | 0.68 | **2.21** | |
| 120 | 0.17 | 0.35 | 0.41 | **1.07** | |
| 140 | 0.11 | 0.22 | 0.28 | **0.66** | |
| 160 | 0.08 | 0.16 | 0.21 | **0.46** | |
| (c) SPI Values for Routes in the hypothetical Scenario at Different Thresholds | | | | | |
| 130 | **5.98** | 1.22 | 1.22 | | |
| 150 | **1.21** | 0.62 | 0.61 | | |
| 170 | **0.53** | 0.37 | 0.36 | | |
| 210 | **0.20** | 0.19 | 0.18 | | |

### 4.4. Comparative Analysis

GARSA is compared with three baseline methods: the A* algorithm, the visibility graph (VG) method [17], and the ship-shore distance (SSD) method [28]. As shown in Figure 12. The A* algorithm and VG method mainly select shorter routes, which pass through the central narrow region of the map, where the dynamic width characteristic is 14 m. As a result, these routes show lower safety level. The SSD method incorporates ship-shore distance as a risk cost function based on the A* algorithm, allowing the route to maintain an approximately 20 m clearance from the shoreline and avoid the narrowest central region. However, SSD mainly improves local clearance and does not fully represent the cumulative spatial constraints along irregular waterways. In such environments, even when the minimum dynamic width exceeds the safety-distance threshold, the overall route risk may remain high due to the cumulative effect of multiple locally narrow segments.

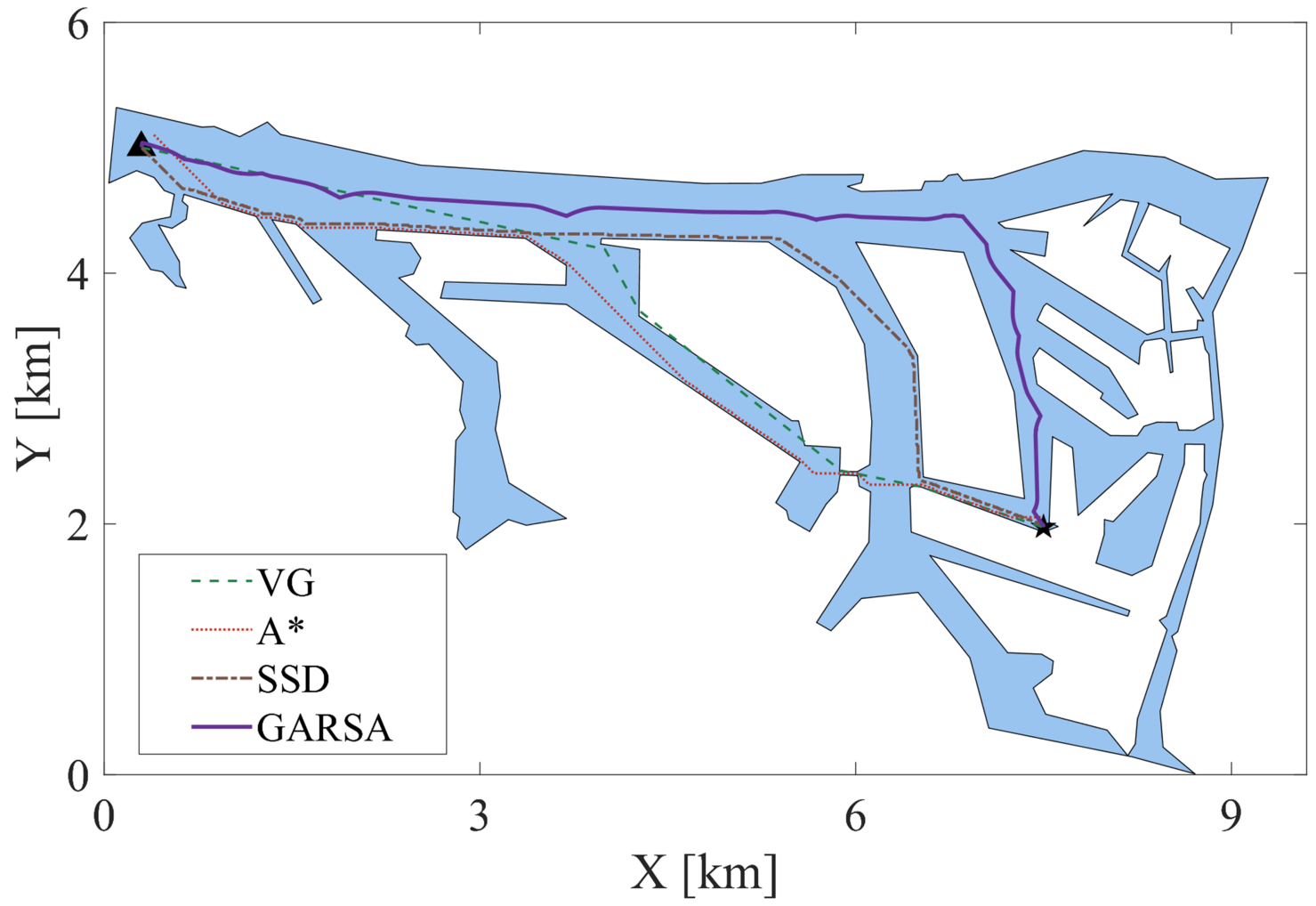

Fig. 12. Illustration of Waterway Decision Comparison in the Port of Hamburg

## 5. Discussion

This study focuses on spatial safety assessment and route decision-making in irregular waterways by representing waterway boundaries with point and line features. A key characteristic of irregular waterways is the non-uniform variation of navigable space along a route. For example, Inbound Route 1 in the Port of Hamburg shows a sharp decrease in the dynamic width characteristic from 173 m to 14 m over a route distance of approximately 170 m. To capture such spatial variation, an iterative computational strategy is used to construct the dynamic width characterization function. This function provides a quantitative representation of waterway spatial constraints and supports safety-oriented route evaluation.

The route decisions indicate that GARSA tends to select routes with more favorable spatial safety characteristics, such as wider navigable space and fewer locally narrow segments. This safety-oriented feature has two main aspects: (i) the evaluation is not primarily determined by route length but width, as SPIs are calculated from the dynamic width characterization function; (ii) the evaluation is not dominated by a single width-based indicator, such as the minimum or average dynamic width characteristic. Instead, SPI accounts for locally narrow segments by amplifying their risk contribution through a logarithmic function and accumulating these contributions along the route. As a result, both the severity and frequency of local narrowing jointly affect the overall safety evaluation.

Previous studies have shown that spatial constraints are important factors affecting vessel safety in restricted waters [15, 27]. However, existing safety assessment methods often represent navigational space through discrete safety levels [20] or focus on local clearance indicators, such as the ship-shore distance [24, 28]. These approaches are useful for identifying local risks, but they may have limited ability to describe the continuous and cumulative spatial variations along irregular waterways. In such environments, route safety is not determined only by the narrowest section or the average navigable width; multiple locally constrained segments may jointly increase the overall risk. The proposed SPI addresses this issue by combining the overall dynamic width characteristic with the accumulated risk contributions of narrow segments, thereby providing a more comprehensive representation of spatial safety in irregular waterways.

This study constructs a quantitative index and a strategic route decision-making framework for spatial safety in irregular waterways. The proposed method is mainly intended for offline or strategic route evaluation at the waterway-layout level, rather than real-time navigation or collision avoidance. In the current framework, water depth constraints, traffic organization requirements, and regulatory compliance are not explicitly modeled. The point-line feature representation used in the case studies is also manually constructed from satellite imagery. Some of these factors can be incorporated as additional constraints depending on their nature. For example, depth-restricted areas can be identified using bathymetric data and converted into non-navigable zones, which can then be represented as additional geometric boundaries in the model. Similarly, traffic organization requirements, such as designated navigation areas or traffic separation schemes, may be introduced by redefining the navigable domain or adding rule-based constraints for route evaluation.

It should be noted that COLREG-related requirements and dynamic obstacles mainly involve vessel interactions and real-time navigation decisions and therefore are beyond the direct scope of the current method. Nevertheless, the routes generated by GARSA can serve as a strategic reference for subsequent local planning or collision-avoidance modules by helping identify route options with more favorable spatial safety characteristics.

## 6. Conclusion

This paper presents a Geometric Analysis-based Route Safety Assessment (GARSA) method for route safety assessment in irregular waterways, prioritizing the safety issue in route decision-making for confined waters. Simulated experiments conducted in Hamburg Port validate GARSA's effectiveness in complex environments, leading to the following conclusions: (i) The proposed method can identify safe routes in narrow, complex, and winding waterways, reflecting a spatial safety-oriented decision strategy. The results are independent of route length and are not influenced by single factors, such as minimum or average dynamic width characteristics; (ii) The dynamic width characterization function captures spatial variations in irregular waterways and provides a quantitative representation of navigable space under geometric boundary constraints, supporting route safety assessment in complex environments; (iii) The Safety Performance Index (SPI) integrates both global and local spatial safety, ensuring that navigational safety is not dominated by single extreme values. Instead, it reflects a balanced assessment of overall channel conditions and locally narrow segments through their combined severity and frequency. The proposed method provides a basis for safety-oriented route decision-making in irregular waterways. Future work is to extend this framework by integrating additional objectives, such as travel distance, time efficiency, and fuel consumption, to support multi-objective route optimization.

**Acknowledgements**

This work was supported in part by the National Natural Science Foundations of China under Grant 52101361.